\documentclass[reprint,pra,groupedaddress]{revtex4-1}
\usepackage{amsmath}
\usepackage{amssymb}
\usepackage{enumitem}
\usepackage{graphicx}
\usepackage{dcolumn}
\usepackage{multirow}
\usepackage{color}
\newcommand{\Hamiltonian}{\mathcal{H}}

\DeclareMathOperator\Ci{Ci}
\DeclareMathOperator\Si{Si}
\DeclareMathOperator\Ciw{Ciw}
\DeclareMathOperator\Siw{Siw}

\DeclareMathOperator{\sgn}{sgn}

\newcommand{\units}[1]{\ensuremath{\mathrm{#1}}}

\newcommand{\ket}[1]{\ensuremath{|{#1}\rangle}}

\newcommand{\red}[1]{\textcolor{black}{#1}}

\newcommand{\beginsupplement}{%
        \setcounter{table}{0}
        \renewcommand{\thetable}{S\arabic{table}}%
        \setcounter{figure}{0}
        \renewcommand{\thefigure}{S\arabic{figure}}%
        \setcounter{equation}{0}
        \renewcommand{\theequation}{S\arabic{equation}}
        \setcounter{section}{0}
        \renewcommand{\thesection}{S\Roman{section}}
        }

\begin{document}


\title{Achieving High-Fidelity Single-Qubit Gates in a Strongly Driven Charge Qubit with $1\!/\!f$ Charge Noise}

\author{Yuan-Chi Yang}
\email[]{yang339@wisc.edu}
\affiliation{Department of Physics, University of Wisconsin-Madison, Madison, Wisconsin, 53706, USA}

\author{S. N. Coppersmith}
\email[]{snc@physics.wisc.edu}
\affiliation{Department of Physics, University of Wisconsin-Madison, Madison, Wisconsin, 53706, USA}

\author{Mark Friesen}
\email[]{friesen@physics.wisc.edu}
\affiliation{Department of Physics, University of Wisconsin-Madison, Madison, Wisconsin, 53706, USA}

\date{\today}

\begin{abstract}
Charge qubits formed in double quantum dots represent quintessential two-level systems that enjoy both ease of control and efficient readout.  
Unfortunately, charge noise can cause rapid decoherence, with typical single-qubit gate fidelities falling below $90\%$.  
Here, we develop analytical methods to study the evolution of strongly driven charge qubits, for general and $1\!/\!f$ charge-noise spectra. 
We show that special pulsing techniques can simultaneously suppress errors due to strong driving and charge noise, yielding single-qubit gates with fidelities above $99.9\%$.   
These results demonstrate that quantum dot charge qubits provide a potential route to high-fidelity quantum computation.
\end{abstract}


\maketitle

\section*{Introduction}
Building high-quality qubits is a key objective in quantum information processing. 
Achieving high-fidelity gates requires both precise control and effective measures to combat decoherence arising from the environment.
Semiconductor based quantum dot charge qubits, for example, suffer from strong coupling to charge noise that causes voltage fluctuations on the control electrodes~\cite{PhysRevLett.105.246804, PhysRevLett.110.146804,2016arXiv161104945T}, which has so far limited gate fidelities to below $90\%$~\cite{Kim2015}. 
To be suitable for scalable quantum computation, the fidelity must be increased to at least $99\%$~\cite{PhysRevA.86.032324}.

\red{
One strategy for achieving higher fidelities is to operate the qubits as fast as possible, for example, by driving them with strong microwaves.
AC driving also mitigates decoherence, by elevating the relevant noise frequencies to the microwave regime, where their power is suppressed~\cite{Wong2016,10.1038/ncomms3337}.
However high-power microwaves can potentially cause detrimental strong-driving effects, including Bloch-Siegert shifts of the resonant frequency~\cite{Bloch1940,Shirley1965,PhysRevB.92.054422} and fast oscillations superimposed on top of Rabi oscillations~\cite{PhysRevA.95.062321}. They can also expose the qubit to new types of decoherence such as dephasing caused by noise-induced variations of the Rabi frequency~\cite{Wong2016,10.1038/ncomms3337}.
While Bloch-Siegert shifts can be accommodated by adjusting the driving frequency or gate time, and the induced decoherence can be suppressed by employing AC sweet spots~\cite{2018arXiv180701310D},
fast oscillations may be difficult to control, resulting in gate errors.}
There are several known approaches for mitigating control errors, including pulse-shaping methods that suppress oscillations by engineering the pulse envelopes~\cite{Motzoi2009, Motzoi2013,PhysRevA.95.062321}.
However, such schemes tend to increase the complexity of the control procedure.

Here we propose an alternative control scheme for strong driving, based on rectangular pulse envelopes engineered to produce nodes in the fast oscillations at the end of a gate operation, thereby minimizing their influence.
We demonstrate our method on a double-quantum-dot charge qubit, showing that high-fidelity gate operations can be achieved in charge qubits under strong driving, even while $1/f$ noise is applied to the double-dot detuning parameter.
\red{
This noise spectrum is particularly interesting because it has both Markovian and non-Markovian components.
By employing both numerical and analytical techniques, we identify specific rotations that synchronize Rabi and fast oscillations, yielding a complete set of single-qubit gates that suppress control errors.
We then propose a protocol for suppressing decoherence caused by charge noise, yielding gates with fidelities higher than $99.9\%$, for typical charge noise magnitudes~\cite{PhysRevLett.105.246804, Wu19082014,PhysRevB.88.075416,2016arXiv161104945T}.
}

\red{
We also develop an analytical formalism based on a cumulant expansion, to accurately describe qubit dynamics in the presence of time-averaged $1/f$ noise.
This formalism allows us explicitly calculate and distinguish between strong driving control errors and decoherence occurring in the weak and strong driving limit.
}

\begin{figure*}[t]
\includegraphics[width=6.5in]{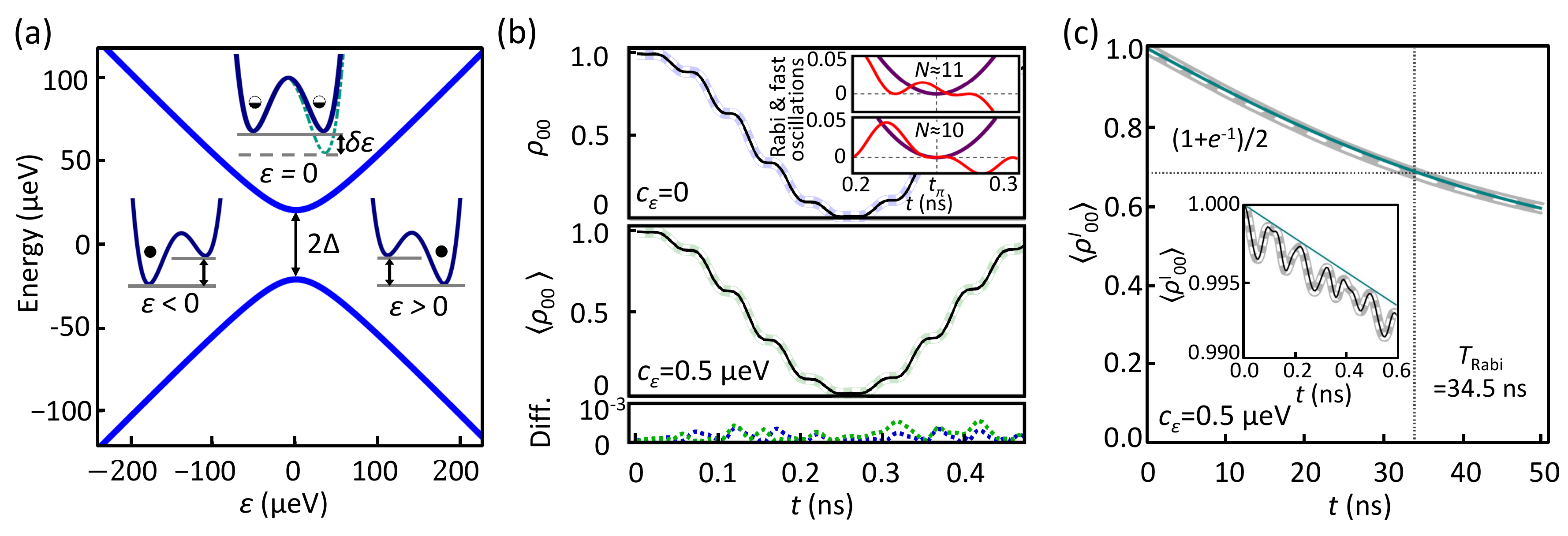}
\caption{
\label{Fig:Fig1}
\red{Strongly driven charge qubit. }
(a) Energy level diagram of a charge qubit. 
The insets depict a double quantum dot in three regimes of detuning, $\varepsilon$. 
(A potential charge noise fluctuation is shown as a dashed line.)
Here, filled circles indicate the position of the excess electron in the ground state, and the barrier between the dots induces a tunnel coupling, $\Delta$.
\red{(b) Time evolution of the density matrix element $\rho_{00}$ in the laboratory frame, including numerical simulations (dashed lines), analytical calculations obtained from Eq.~(\ref{Eq:AnalyticFormula2nd}) (solid black lines), and their differences (dotted lines). (Here, brackets $\langle\cdot\rangle$ denote a noise average.)
In all cases, we use $\{\varepsilon, \Delta, A_{\varepsilon}\}/h = \{0, 5 ,4\} \,\units{GHz}$, $\phi = \pi/4$ and initial state $\ket{0}$.
We assume the charge noise follows the $1\!/\!f$ spectrum of Eq.~(\ref{Eq:Spectrum One-Over-f}), with frequency cutoffs $\omega_{l}/2\pi = 0.193 \,\units{MHz}$ and $\omega_{h}/2\pi = 80.8 \,\units{GHz}$, and noise amplitudes $c_\varepsilon$ as indicated.
The insets of (b) show blow-ups of the evolution near the end of a $\pi$-rotation ($t=t_\pi$), decomposed into their Rabi (dark purple) and fast-oscillation components (red).
The oscillations are synchronized at $t_\pi$ when $N=(2\theta\tilde \omega_\text{res})/(\pi \Omega)$ is an even integer, resulting in high-fidelity gates.
(The main panels also use $N=10$.)
Charge noise causes a slight decay of $\langle\rho_{00}\rangle$ at the end of the simulation period ($2t_\pi\simeq 0.5$~ns), which can be observed more clearly at long times in (c).
(c) Time evolution of the density matrix element $\rho_{00}^I$ in the interaction frame, including numerical simulations (dashed gray line) and the simple asymptotic expression from Eq.~(\ref{Eq:AymptoticFormula}) (solid cyan line) with corrections to $K_{\varphi}$ up to $O[\gamma^2]$ and corrections to $K_\text{M}$ and $K_{\text{nMn}\varphi}$ up to $O[\gamma]$, as discussed in Supplementary Sec. S3.
The inset shows a short-time blow-up in the interaction frame; it further includes our full analytical calculations obtained from Eq.~(\ref{Eq:AnalyticFormula2nd}) (solid black line).}
}
\end{figure*}

\begin{figure}[t]
\includegraphics[width=2.4in]{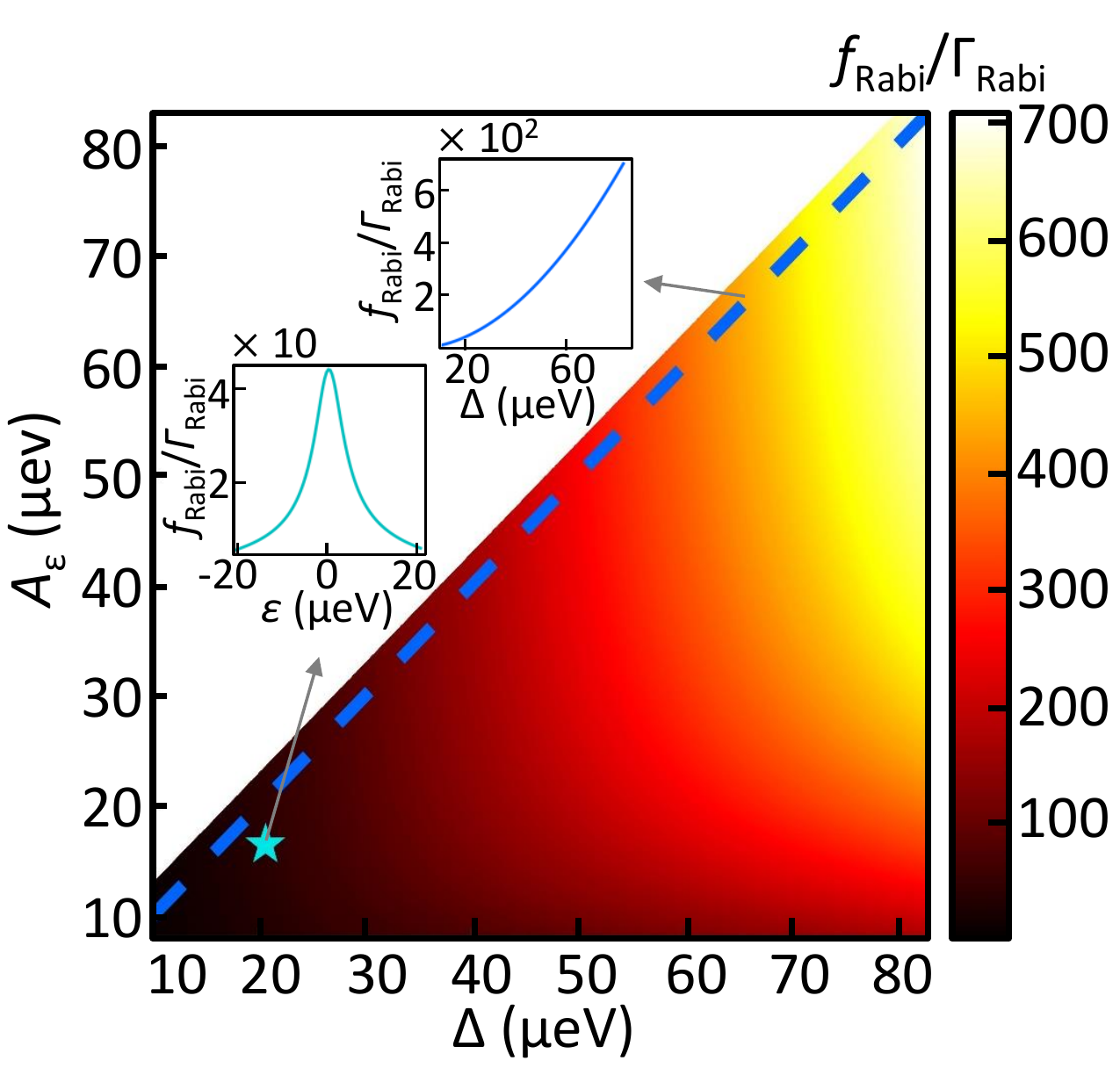}
\caption{
\label{Fig:Fig2}
\red{Figure of merit (FOM) of gates of a strongly driven charge qubit.}
\red{Analytical calculations of the FOM} $f_\text{Rabi}/\Gamma_\text{Rabi}$ of a strongly driven charge qubit, as a function of the tunnel coupling $\Delta$ and driving amplitude $A_{\epsilon}$, at $\varepsilon=0$ and $\phi = \pi/4$, based on the asymptotic formula in Eq.~(\ref{Eq:AymptoticFormula}), with corrections to $K_{\varphi}$ up to $O[\gamma^2]$ and corrections to $K_\text{M}$ and $K_{\text{nMn}\varphi}$ up to $O[\gamma]$, as discussed in Supplementary Sec.~S3.
Here, $f_\text{Rabi}$ is the Rabi frequency, $\Gamma_\text{Rabi}$ is the Rabi decay rate, and the $1\!/\!f$ noise spectrum is given in Eq.~(\ref{Eq:Spectrum One-Over-f}), with $c_{\varepsilon} = 0.5 \, \units{\mu eV}$, $\omega_l/2\pi = 1 \, \units{Hz}$, and $\omega_h/2\pi = 100 \, \units{THz}$. 
The upper inset shows a line-cut along $A_{\varepsilon} = \Delta$ in the main figure (blue dashed line), revealing a FOM as high as 700.
In the lower inset, we fix $\Delta/h = 5 \,\units{GHz}$ and $A_{\varepsilon}/h = 4 \,\units{GHz}$ (cyan star), but allow $\varepsilon$ to vary, confirming that the FOM is maximized at the sweet spot, $\varepsilon=0$, where the qubit is first-order insensitive to detuning noise.
}
\end{figure}

\section*{Results}
\textbf{Noise-free evolution.}
The basis states of a double quantum dot charge qubit, $\ket{L}$ and $\ket{R}$, represent the localized positions of an excess charge in the left or right dot, as indicated in Fig.~\ref{Fig:Fig1}(a)~\cite{PhysRevLett.91.226804, PhysRevLett.95.090502, Kim2015}. 
We consider ac gating of a single qubit, with the Hamiltonian
$\Hamiltonian_\text{sys}$=$\Hamiltonian_q + \Hamiltonian_\text{ac}$, where $\Hamiltonian_q$=$-(\varepsilon/2) \sigma_x - \Delta \sigma_z$,  the $\sigma_i$ are Pauli matrices, $\varepsilon$ is the detuning parameter (defined as the energy difference between the two dots), and $\Delta$ is the tunnel coupling between the dots.
Here we have expressed $\Hamiltonian_\text{sys}$ in the eigenbasis $\{|0\rangle = (|L\rangle-|R\rangle)/\sqrt{2},|1\rangle = (|L\rangle+|R\rangle)/\sqrt{2}\}$, corresponding to the charge qubit ``sweet spot" $\varepsilon$=0, where it is first-order insensitive to electrical noise~\cite{Kim2015}.
Unless otherwise noted, we assume that the nominal operating point is $\varepsilon$=0 throughout the remainder of this work.
When a microwave signal is applied to $\varepsilon$, the driving Hamiltonian is given by $\Hamiltonian_\text{ac} =(A_{\varepsilon}/2)\sigma_x \cos(\omega_d t+\phi)$, where $A_{\varepsilon}$ is the driving amplitude, $\omega_d$ is its angular frequency, and $\phi$ is the phase at time $t$=0, when the drive is initiated.
 

First, we follow Ref.~\cite{PhysRevA.95.062321} and obtain exact solutions for strongly-driven qubits in the absence of noise, up to arbitrary order in the strong-driving parameter $\gamma$=$A_{\varepsilon}/(16\Delta)$. 
Expanding the time-evolution operator order-by-order as $U_0(t)$=$\sum_{n=0}^\infty \gamma^n U_0^{(n)}$
, we obtain 
\begin{equation}
\label{Eq:U00}
U_0^{(0)}\!(t) \!= \!
\begin{pmatrix}
e^{i\tilde{\omega}_\text{res} t} \cos(\Omega t/2) &  \!\text{-}i e^{i(\tilde{\omega}_\text{res} t+\phi)}\sin(\Omega t/2)\\
 \text{-}i e^{-i \phi} \sin(\Omega t/2) &  \cos(\Omega t/2)
\end{pmatrix} \!. 
\end{equation}
[Higher-order terms are provided in \red{Supplementary Section S2, Eq.~(S5).}]
Here, we consider only resonant driving, $\omega_d$=$\tilde\omega_\text{res}$, where $\hbar \tilde{\omega}_\text{res}$=$2 \Delta (1+4 \gamma^2)$ is the renormalized resonant angular frequency, including Bloch-Siegert corrections, and $\hbar \Omega$=$A_{\varepsilon}(1+\gamma^2)/2$ is the renormalized Rabi angular frequency. 

In the rotating frame defined by $\Hamiltonian_\text{rot}$=$U_\text{rot}^{\dagger} \Hamiltonian_\text{sys} U_\text{rot} - i \hbar U_\text{rot}^{\dagger} (d/dt) U_\text{rot} $, with $U_\text{rot}$=$\text{diag}[e^{i\omega_\text{d} t/2},e^{-i\omega_\text{d} t/2}]$, the ideal evolution term $U_0^{(0)}$ generates smooth, sinusoidal, Rabi oscillations, corresponding to rotations about the $(\cos\phi,-\sin\phi,0)$ axis of the Bloch sphere.
The $U_0^{(1)}$ term represents the dominant fast oscillations associated with strong driving, with amplitude $\sim$$\gamma$. 
Both drive components can be observed in the top panel of Fig.~\ref{Fig:Fig1}(b), where we show that our analytical results (shown here up to $O[\gamma^2]$) agree well with the results of numerical simulations of the full Hamiltonian. 

Fast oscillations can cause gate infidelity. 
For example, if we consider an $R_\theta(\phi)$ rotation of angle $\theta$ about the $(\cos\phi,-\sin\phi,0)$ axis, the fast oscillations may prevent the density matrix element $\rho_{00}$ from reaching $0$ at the end of a gate period, $t_{\pi}$. 
We see this more clearly by plotting the fast and Rabi oscillation components separately in the top inset of Fig.~\ref{Fig:Fig1}(b).
On the other hand, we may adjust the pulse parameters $A_\varepsilon$ and $\phi$ to synchronize the fast oscillations with the slower Rabi oscillations, as shown in the bottom inset, to obtain an $R_\theta(\phi)$ gate with much higher fidelity. 

\red{
We characterize the infidelity arising from the fast oscillations by computing the process fidelity $F_\theta(\phi)$, defined by comparing the ideal evolution operator $U_0^{(0)}$ to the actual evolution $U_0$, for $R_\theta(\phi)$ in the rotating frame. [see Eq.~(S2) of Supplementary Section S1 for a precise definition of the process fidelity.]}
We find that specific $A_\varepsilon$'s give rotations with perfect fidelity when $\theta=\pi,2\pi,3\pi,\dots$ for any $\phi$.
More importantly, when $\phi=\pi/4,3\pi/4,5\pi/4,\dots$, we obtain 
\begin{multline}
1- F_{\theta}(\phi) = 2 \gamma^2 [1 - \cos(2 \theta \tilde{\omega}_\text{res}/\Omega )] \\
+4\gamma^3\sin\theta\sin(2 \theta \tilde{\omega}_\text{res}/\Omega ) +O[\gamma^4] ,
\label{Eq:FidelityFormulaNNoise}
\end{multline} 
where, again, $\gamma = A_{\varepsilon}/(16\Delta)$. For these values of $\phi$, the infidelity due to strong-driving errors is bounded above by $\sim$$4\gamma^2$.
Moreover, the oscillations are synchronized, yielding perfect fidelity (up to $O[\gamma^4]$), when $2 \theta \tilde{\omega}_\text{res}/\Omega$=$N\pi$, with $N$ an even integer.
Since this condition can be met for a continuous range of $\theta$ by adjusting $\Omega$ (i.e., $A_\varepsilon$), and since $\phi=\pi/4$ and $3\pi/4$ represent orthogonal rotation axes, the rotations $\{R_\theta(\pi/4),R_\theta(3\pi/4)\}$ therefore generate a complete set of high-fidelity single-qubit gates.
Additional phase control is provided by adjusting the waiting time between ac pulses.
Unless otherwise noted, we set $\phi=\pi/4$ for the remainder of our analysis.

\textbf{Charge noise.}
We introduce charge noise into our analysis through the Hamiltonian term $\Hamiltonian_n= h_n \delta \varepsilon(t)$, where $\delta \varepsilon(t)$ is a random variable affecting the detuning parameter and $h_n = -\sigma_x/2$ is referred to as the noise matrix~\cite{PhysRevLett.105.246804, Wu19082014,PhysRevB.88.075416,2016arXiv161104945T}. 
The noise is characterized in terms of its time correlation function $S(t_1-t_2) = \langle \delta\varepsilon(t_1) \delta\varepsilon(t_2) \rangle$, where the brackets denote an average over noise realizations, and the corresponding noise power spectrum is $\tilde{S}(\omega) = \int_{-\infty}^{\infty} dt\,e^{i \omega t} S(t)$ \cite{RevModPhys.82.1155}.
Although we obtain analytical solutions for generic noise spectra in Supplementary Sec.~S3, below we focus on $1\!/\!f$ noise, including in our simulations, due to its relevance for charge noise in semiconducting devices~\cite{nphys2688,RevModPhys.86.361}:
\begin{equation}
\label{Eq:Spectrum One-Over-f}
\tilde{S}(\omega) =  \left\{
  \begin{array}{cl}
   c_{\varepsilon}^2 \frac{2 \pi}{|\omega|} &  (\omega_l \leq |\omega | \leq \omega_h)\\
    0 &  (\text{otherwise}) 
  \end{array}
\right. ,
\end{equation}
where $c_\varepsilon$ is related to the standard deviation of the detuning noise, $\sigma_{\varepsilon}$, via $\sigma_{\varepsilon} = c_{\varepsilon}[2 \ln(\sqrt{2\pi}c_{\varepsilon}/\hbar\omega_l)]^{1/2}$ \cite{MAKHLIN2004315,Wong2016}, and $\omega_l$ ($\omega_h$) are low (high) cutoff angular frequencies.
\red{We note that all frequencies relevant for qubit operation occur between these two cutoffs, so that the decoherence includes both Markovian and non-Markovain contributions.}

We now present numerical simulations of a strongly driven, noisy charge qubit.
A typical result is shown in the middle panel of Fig.~\ref{Fig:Fig1}(b), where the suppression of Rabi oscillations is a direct consequence of the charge noise.
To differentiate the effects of decoherence from those arising from strong driving, we present the same results in an interaction frame defined by $U_0$, $\rho^I = U_0^{\dagger} \rho U_0$, in which the fast oscillations due to strong driving are not observed. 
Figure~\ref{Fig:Fig1}(c) shows the resulting long-time decay of the density matrix, while the inset shows the short-time behavior on an expanded scale.
Note that the fast oscillations observed here do not arise directly from strong driving, but rather from non-Markovian noise terms, as discussed below. 
 
\textbf{Analytical solutions, with charge noise.}
Several theoretical techniques have been applied to noisy, driven two-level systems, including master equations~\cite{doi:10.1063/1.468844,PhysRevB.67.155104,10.1038/ncomms3337,PhysRevA.90.022118,PhysRevB.72.134519,Wong2016,Slitcher1990}, dissipative Lander-Zener-St\"{u}ckelberg interferometry~\cite{SHEVCHENKO20101,PhysRevB.93.064521}, and treatments of spin-Boson systems~\cite{Goorden2005, GRIFONI1998229,PhysRevE.61.R4687}. 
In Supplementary Sec.~S3, we solve the dynamical equation in the interaction frame via a cumulant expansion~\cite{doi:10.1143/JPSJ.17.1100, doi:10.1063/1.1703941}, truncated at $O[(\delta \varepsilon/\hbar \Omega)^2]$.
The time evolution can be written in the form $\mathbf{r}^I(t) = \exp [K(t)] \mathbf{r}^I(0)$ by expressing $\langle \rho^I \rangle = 1/2 (I_2 +r^I_x \sigma_x + r^I_y \sigma_y + r^I_z \sigma_z)$. 
Here, $I_2$ is the $2 \times 2$ identity matrix, $\mathbf{r}^I = (r^I_x, r^I_y , r^I_z)$ is the Bloch vector, and $K(t)$ is a $3 \times 3$ evolution matrix, given by
\begin{eqnarray}
\label{Eq:AnalyticFormula2nd}
&&[K(t)]_{ij}  \\ 
&&= -\frac{4}{\hbar^2} \sum\limits_{\omega_1, \omega_2}
[\red{-}\alpha_{j,\omega_1}\alpha_{i,\omega_2}  \red{+} \delta_{ij} \sum\limits_{k=1}^3\alpha_{k,\omega_1}\alpha_{k,\omega_2} ] I(t,\omega_1,\omega_2), \nonumber
\end{eqnarray}
where we have expand the noise matrix in the interaction frame into Fourier components $h_n^I(t) \equiv U_0^{\dagger} h_n U_0 = \sum_{i,\omega}\alpha_{i,\omega} e^{i\omega t}$, and defined $I(t,\omega_1,\omega_2)  \equiv \int_0^t dt_1 \int_0^{t_1} dt_2 e^{i\omega_1 t_1} e^{i\omega_2 t_2} S(t_1-t_2)$. 
\red{Since $U_0$ can be expressed order-by-order in $\gamma$, the same is also true of $\alpha_{i,\omega}$, allowing us to distinguish the effects arising in the weak-drive limit, $O[\gamma^0]$, from the strong-driving limit, $O[\gamma^n]$, for $n\geq 1$.}
The accuracy of this cumulant approach is evident in Figs.~\ref{Fig:Fig1}(b) and \ref{Fig:Fig1}(c), where the theoretical results (\red{solid black line}) are seen to capture all the fine structure of the simulations. 
Indeed, the bottom panel of Fig.~\ref{Fig:Fig1}(b) indicates that the analytical and numerical solutions differ by $< 10^{-3}$ over the entire range plotted.


\begin{figure*}[t]
\includegraphics[width=6.in]{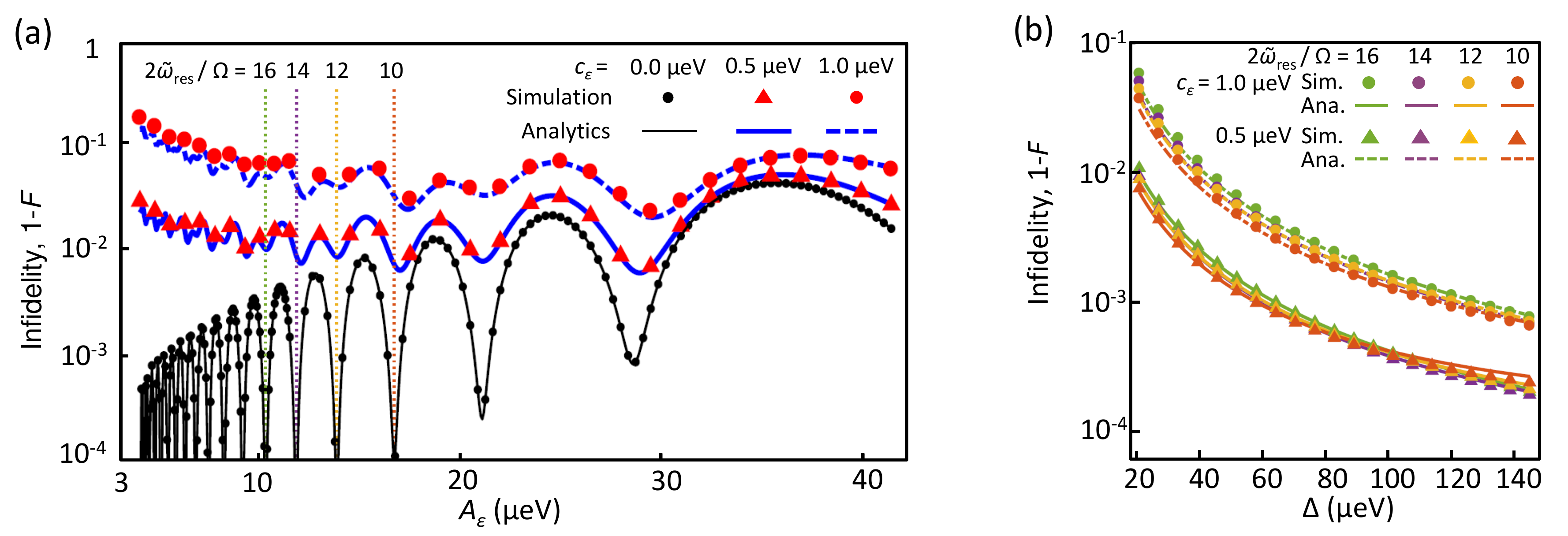}
\caption{
\label{Fig:Fig3}
\red{Characterization of the fidelity.}
Dependence of the infidelity $1$-$F_{\pi}(\pi/4)$ of a strongly driven charge qubit on (a) the driving amplitude $A_\varepsilon$, with $\Delta/h = 5 \, \units{GHz}$, and on (b) the tunnel coupling $\Delta$, with $N$=$(2 \theta \tilde{\omega}_\text{res})/(\pi \Omega)$=10, 12, 14, 16 held constant. 
Here, $\varepsilon = 0$, and the charge noise is given by Eq.~(\ref{Eq:Spectrum One-Over-f}), with $\omega_l/2\pi = 1 \, \units{Hz}$, $\omega_h/2 \pi = 256 \, \units{GHz}$, and the values of $c_\varepsilon$ are indicated.
Note that low frequencies ($< 0.3 \, \units{MHz}$) are approximated as quasistatic in these simulations, while the analytical results are obtained from Eq.~(\ref{Eq:AnalyticFormula2nd}), up to $O[\gamma^3]$ [or $O[\gamma^4]$ for $c_\varepsilon=0$, see Supplementary Eq.~(S12)].
In (b), the infidelities are computed at the ``dips" indicated in (a).
}
\end{figure*}

The physics of noise-averaged qubit dynamics is encoded in $K(t)$, which can be decomposed into a sum of Markovian terms $K_\text{M}$, and non-Markovian terms.
The latter may be further divided into pure-dephasing terms $K_{\varphi}$~\cite{PhysRevLett.88.228304,PhysRevLett.93.267007, PhysRevB.72.134519}, and non-Markovian-non-dephasing terms $K_{\text{nMn}\varphi}$. 
Pure-dephasing terms are conventionally associated with the integral $I(t,\omega_1\text{=}0,\omega_2$=$0) \sim t^2\ln(1/\omega_l t)$~\cite{PhysRevLett.88.228304,PhysRevLett.93.267007, PhysRevB.72.134519}.
However, since $K$ is defined in a rotating frame, ``pure dephasing" has a different meaning than in the laboratory frame~\cite{10.1038/ncomms3337, PhysRevB.72.134519}: here, the leading order contributions to $K_{\varphi}$ are proportional to $\gamma^2$, and are therefore attributed to strong driving.
Markovian terms are associated with the integral $\text{Re}[I(t,\omega,-\omega)]$, corresponding to short correlation times [see Supplementary Eq.~(S34)], and exponential decay ($\sim$$e^{-\Gamma t}$).
The dominant non-Markovian-non-dephasing terms are associated with the integral $\text{Im}[I(t,\omega,-\omega)]$], yielding slow oscillations in the rotating frame, as well as the fast oscillations in the inset of Fig.~\ref{Fig:Fig1}(c).
Since it is common in the literature to treat the dephasing and depolarizing channels separately~\cite{PhysRevB.72.134519},
it is significant that our method encompasses both phenomena (and other behavior, including $K_{\text{nMn}\varphi}$) within a common framework, allowing us to compare and contrast their effects. 

\textbf{Asymptotic solutions.}
We can compute the asymptotic behavior of $\mathbf{r}^I(t)$ analytically, using the cumulant expansion.
In the limit $t \gg 1/\omega$, where $\omega$ is any characteristic qubit frequency, many terms drop out, yielding the leading-order solution in $\gamma$:
\begin{equation}
\label{Eq:AymptoticFormula}
\mathbf{r}^I(t) \!=\! \frac{e^{ - \Gamma_{z} t}}{\sqrt{2}}[\sin(\Gamma_{\text{nMn}\varphi} t),\sin(\Gamma_{\text{nMn}\varphi} t),\!\sqrt{2}\cos(\Gamma_{\text{nMn}\varphi} t)],
\end{equation}
for the initial state $|0\rangle$.
Here, the decoherence is dominated by the integral
$I(t,\omega,-\omega)\, \approx \, \tilde S(\omega) t/2 + \tilde{S}_\text{imag}(\omega) t$, whose imaginary part is given by $\tilde{S}_\text{imag}(\omega) \equiv c_{\varepsilon}^2(2 i/\omega) \ln|\omega/\omega_l|$.
The real part describes exponential decay, giving the Markovian decoherence rate for driven evolution,
$\Gamma_{z}$=$
(1/16\hbar^2)\,[4 \tilde{S}(\tilde{\omega}_\text{res})$+$\tilde{S}(\tilde{\omega}_\text{res}$+$\Omega)$+$\tilde{S}(\tilde{\omega}_\text{res}$$-$$\Omega)] $ as observed in Fig.~\ref{Fig:Fig1}(c), which can also be derived from Bloch-Redfield theory~\cite{PhysRevB.72.134519, Slitcher1990}.
The imaginary part corresponds to a non-Markovian-non-dephasing noise-induced rotation with frequency $\Gamma_{\text{nMn}\varphi}=(-i/8\hbar^2)\, [(\tilde{S}_\text{imag}(-\tilde{\omega}_\text{res} $+$\Omega)$+$\tilde{S}_\text{imag}(\tilde{\omega}_\text{res}$+$\Omega))]$, originating from the integrated low-frequency (``quasistatic") portion of the noise spectrum, $[\int_{\omega_l}^{\omega} d \omega^{\prime}S(\omega^{\prime})/\pi ]^{1/2}=c_{\varepsilon}[2 \ln(\omega/\omega_l )]^{1/2}$.
\red{Besides these lowest-order results, which are the only important terms under weak driving, we can also compute higher-order corrections to these terms that become important under strong driving.
Such high-order results are presented in Supplementary Section~S3.}

\red{We can define a figure of merit (FOM), $f_\text{Rabi}/\Gamma_\text{Rabi}$, corresponding to the number of coherent $R_{2\pi}(\pi/4)$ rotations within a Rabi decay period $T_\text{Rabi} = 1/\Gamma_\text{Rabi}$ (not including strong-driving control errors), where the Rabi decay rate $\Gamma_\text{Rabi}$ is determined from Eq.~(\ref{Eq:AymptoticFormula}) such that $\langle \rho^I_{00}\left(t =1/\Gamma_\text{Rabi}\right)\rangle \equiv (1 + e^{-1})/2$.}
By exploring a range of control parameters in Fig.~\ref{Fig:Fig2}, we first confirm that the FOM is strongly enhanced at the sweet spot $\varepsilon = 0$ (lower inset). 
Increasing the tunnel coupling $\Delta$ and the driving amplitude $A_{\varepsilon}$ both enhance the FOM, as shown in the main panel.
By increasing $\Delta$ and $A_{\varepsilon}$ simultaneously, as shown in the upper inset, we find that the FOM can exceed $700$ for a physically realistic charge noise amplitude of $c_{\varepsilon} = 0.5 \, \units{\mu eV}$ ($\sigma_{\varepsilon} = 3.12 \,\units{\mu eV}$)~\cite{PhysRevLett.105.246804, Wu19082014,PhysRevB.88.075416,2016arXiv161104945T}.

\textbf{$\mathbf{R_{\pi}(\pi/4)}$ gate fidelity.}
We now compute process fidelities for $R_{\pi}(\pi/4)$ gates, using the $\chi$-matrix method described in Supplementary Sec.~S1.
Control errors due to strong driving are investigated by considering $U_0^{(0)}$ as the ideal evolution.
The results of both numerical and analytical calculations are shown in Fig.~\ref{Fig:Fig3}. 
For no noise ($c_{\varepsilon}$=0), the simulations are essentially identical to Eq.~(\ref{Eq:FidelityFormulaNNoise}), revealing ``dips" of low infidelity, enabled by synchronized oscillations.
For $\phi=\pi/4$, the dip minima are proportional to $\gamma^4$, while their widths are proportional to $\gamma^2$ [see Supplementary Eq.~(S13)], suggesting potential benefits of working at large $A_\varepsilon$$\propto$$\gamma$.
\red{
As $c_{\varepsilon}$ increases, the infidelity also grows, including both Markovian ($K_\text{M}$), and non-Markovian contributions ($K_{\varphi}$ and $K_{\text{nMn}\varphi}$).
Initially, the envelope of the infidelity oscillations decreases with $A_\varepsilon$, because fast gates have less time to be affected by noise; it then increases, due to the combinantion of strong-driving effects and the decoherence induced by strong driving.}
For smaller $A_\varepsilon$, the simulations deviate slightly from the analytical results when the high-order noise terms become non-negligible.
In all cases, the infidelity is locally minimized when $A_{\varepsilon}$ is positioned at a dip.

For the noise levels considered in Fig.~\ref{Fig:Fig3}(a), which are consistent with recent experiments~\cite{PhysRevLett.105.246804, Wu19082014,PhysRevB.88.075416,2016arXiv161104945T}, we obtain $F_\text{max}$$\lesssim$$99\%$, which is insufficient for achieving high-fidelity gates.
However, the following procedure can be used to suppress both control errors and decoherence.
First, $A_\varepsilon$ is tuned to a dip.
Then, $\Delta$ and $A_\varepsilon$ are simultaneously increased while holding $\gamma=A_{\varepsilon}/16\Delta$ (and thus $N$) fixed.
In this way, we remain in a dip, while increasing the gate speed to suppress noise effects.
The results are shown in Fig.~\ref{Fig:Fig3}(b).
Here, when $c_{\varepsilon} = 1 \, \units{\mu eV}$ ($\sigma_{\varepsilon} = 6.36 \,\units{\mu eV}$), we obtain fidelities $>$$99\%$ when $\Delta$$>$$40\, \units{\mu eV}$, and $>$$99.9\%$ when $\Delta$$>$$120\, \units{\mu eV}$.
The corresponding qubit frequencies, $2\Delta/h $=$29.3\, \units{GHz}$ and $58.0\, \units{GHz}$, are comparable to the qubit frequency of the quantum dot spin qubit in Ref.~\cite{2018arXiv180301609C}, and the Rabi frequencies $\simeq 4\Delta/h N$ are generally lower.
\red{We note that this protocol is applicable for any rotation angle $\theta$, as shown in Supplementary Section S5.}

\section*{Discussion}
We have developed a new scheme for effectively harnessing strong driving to perform high-fidelity gates in quantum double-dot charge qubits, even in the presence of realistic $1/f$ noise.
Our protocol, and our analytical formalism, are both applicable to other solid-state systems, including superconducting flux qubits~\cite{PhysRevLett.115.133601,PhysRevA.94.032323} and quantum-dot singlet-triplet qubits~\cite{Wu19082014,RN36,doi:10.1063/1.4810911}, and can be extended to systems with multiple levels, including quantum-dot hybrid qubits~\cite{Shi2012,Koh2012,KimShiSimmonsEtAl2014,Ferraro2014,Cao2016,2016arXiv161104945T,PhysRevA.95.062321} and charge-quadrupole qubits~\cite{ncomms15923, PhysRevB.95.241307}. 
\red{
Phonon-induced decoherence can be also analyzed in this formalism, after first averaging the phonons over the corresponding thermal distribution~\cite{doi:10.1063/1.1703941,PhysRevB.89.085410,Kornich2018phononassisted,1367-2630-20-10-103048}.
However, the effectiveness of the protocol may be reduced compared to case of charge noise, since the power spectral density of phonons typically increases with the frequency.
}

A possible challenge for implementing this proposal is the requirement of large tunnel couplings, which could results in fast qubits that are difficult to control.
However, by employing high-order synchronized oscillations (e.g. $N$$\sim$10), we can reduce gate speeds to be compatible with current experiments. 
Improvements in ac control technology and materials with lower charge noise can also mitigate the technical challenges.
On the other hand, the phonon-induced relaxation rate increases strongly with tunnel coupling~\cite{PhysRevLett.111.046801, PhysRevLett.95.106801, RN41}, which will set an upper bound on the qubit coherence.
Moving forward, we note that the phase, $\phi$, represents an important control knob in our proposal, and can be viewed as a simple pulse-shaping tool.
In future work, it should be possible to combine the methods described here with conventional pulse shaping techniques, which would be expected to further improve the gate fidelities~\cite{PhysRevA.95.062321, Motzoi2013, Motzoi2009}.


\section*{Methods}
\textbf{Numerical Simulation.}
We numerically simulate the Schrodinger equation of a strongly driven, noisy charge qubit,  $i \hbar \, d\rho/dt=[\Hamiltonian_\text{sys} + \Hamiltonian_\text{n}, \rho]$, where the time sequences for $\delta \varepsilon(t)$ are obtained by generating a white noise sequence, then scaling its Fourier transform by an appropriate spectral function~\cite{Kawakami18102016}, such as Eq.~(\ref{Eq:Spectrum One-Over-f}).
We then average the density matrix $\langle \rho(t) \rangle$ over noise realizations. 
Details of these procedures are provided in Supplementary Sec.~S4.

\textbf{Analytical Formalism.}
We analytically solve the dynamical equation
$i \hbar \, d\rho^I/dt=\delta \varepsilon(t) \mathcal{L} \rho^I $ in the interaction frame,
where $\mathcal{L} \rho^I \equiv [h_n^I,\rho^I]$ and $h_n^I = U_0^{\dagger} h_n U_0$.  
We then average over the noise via a cumulant expansion~\cite{doi:10.1143/JPSJ.17.1100, doi:10.1063/1.1703941},
truncated at $O[(\delta \varepsilon/\hbar \Omega)^2]$, yielding
$\langle \rho^I(t) \rangle =\exp [ -\frac{1}{\hbar^2}  \int_0^t dt_1 \int_0^{t_1} dt_2 \mathcal{L}(t_1)  \mathcal{L}(t_2) S(t_1-t_2)  ] \rho^I(0)$,
where we have assumed that the noise is stationary, with zero mean ($\langle \delta \varepsilon\rangle = 0$). 
Details of the calculations are provided in Supplementary Sec.~S3.

\red{
\section*{Data availability}
The data and numerical codes that support the findings of this study are available from the
corresponding author upon reasonable request.
}

\section*{Acknowledgement}
We thank M. Eriksson for helpful discussions.
Y.Y. was supported by a Jeff and Lily Chen Distinguished Graduate Fellowship. 
This work was also supported in part by ARO (W911NF-12-1-0607, W911NF-17-1-0274) and the Vannevar Bush
Faculty Fellowship program sponsored by the Basic Research Office of the Assistant Secretary of Defense for
Research and Engineering and funded by the Office of Naval Research through grant N00014-15-1-0029. The views and conclusions contained in this document are those of the authors and should not be interpreted as representing the official policies, either expressed or implied, of the Army Research Office (ARO), or the U.S. Government. The U.S. Government is authorized to reproduce and distribute reprints for Government purposes notwithstanding any copyright notation herein.

\red{
\section*{Competing Interests}
The authors declare that there are no competing interests.
}

\section*{Author Contributions}
Y.Y. performed numerical simulations and analytical calculations. Y.Y., S.N.C., and M.F. analyzed the results and prepared the manuscript. Work was carried out under supervison of S.N.C. and M.F.

\beginsupplement
\begin{widetext}
\section*{Supplementary Information}

In these Supplemental Materials, in Sec.~\ref{section:ProcessFidelity}, we provide the definition of the process fidelity, and in Sec.~\ref{section:fidelityformula} we present a brief derivation of the fidelity formula, Eq.~(2) in the main text.
In Sec~\ref{section:AsymptoticFormula}, we sketch the derivation of the analytic formula for the qubit time evolution for general noise spectra, and derive the asymptotic formulas for Markovian noise, quasi-static noise, and $1/f$ noise. 
Sec.~\ref{section:simulation} provides the details of the simulations shown in the main text, including the method used to generate the noise realizations and a method used to speed up the simulations. 
In Sec~\ref{section:RthetaGates}, we show that high-fidelity rotations about an axis tilted away from $X$-axis by $\phi=\pi/4$ by an arbitrary angle $\theta$, $R_{\theta}(\phi=\pi/4)$, can be achieved using a method similar to the one discussed in the main text.

\section{Process Fidelity}
\label{section:ProcessFidelity}
Following Ref.~\cite{ChuangBook}, a generic quantum process $\mathcal{E}$ on a two dimensional Hilbert space can be expressed as
\begin{equation}
\mathcal{E}(\rho_0) = \sum\limits_{m,n} E_m \rho_0 E_n ^{\dagger} \chi_{mn},
\end{equation}
where $\rho_0$ is the initial-state density matrix, $\{E_m\} = \{I, \sigma_x, -i \sigma_y, \sigma_z\}$ is a basis for the vector space of $2\times 2$ matrices, and $\chi_{mn}$ is a $4\times 4$ process matrix, commonly referred to as the chi matrix.

The process fidelity is then defined as 
\begin{equation}
\label{Eq:ProcessFidelity}
F = \text{Tr} [\chi_\text{sys} \chi_\text{ideal}], 
\end{equation}
where $\chi_\text{sys}$ is the process matrix for the actual physical evolution, including strong-driving effects, and $\chi_\text{ideal}$ is the process matrix for the ideal rotation.

\section{Derivation of Fidelity Formula In The Absence of Charge Noise}
\label{section:fidelityformula}
We now evaluate Eq.~(\ref{Eq:ProcessFidelity}) in the absence of charge noise, including only strong-driving effects.
Additional effects due to charge noise are presented in Sec.~\ref{section:AsymptoticFormula}.

For a resonantly driven charge qubit operated at the charge anti-crossing ($\varepsilon = 0$) with only detuning drive ($A_{\Delta} = 0$), the Hamiltonian in the eigenbasis of the qubit is given by
\begin{equation}
\Hamiltonian_\text{sys} = -\Delta \sigma_z +(A_{\varepsilon}/2)\sigma_x \cos(\tilde{\omega}_\text{res} t + \phi), 
\end{equation}
where $\tilde{\omega}_\text{res}$ is the resonant angular frequency.
The evolution operation of the system in the absence of noise, $\mathcal{E}_\text{sys}(\rho_0) = U_0 \rho_0 U_0^{\dagger}$, can be obtained analytically up to arbitrary order in the strong-driving parameter $\gamma \equiv  A_{\varepsilon}/ (16\Delta)$ as $U_0(t)$=$\sum_{n=0}^\infty \gamma^n U_0^{(n)}$.
The ideal unitary evolution of an $R_{\theta}(\phi)$ rotation corresponds to the $O[\gamma^0]$ terms in this expansion, which can also be obtained from the rotating wave approximation, giving
\begin{equation}
\label{Eq:U00}
U_\text{ideal} = U_0^{(0)} =
\begin{pmatrix}
e^{i\tilde{\omega}_\text{res} t_{\theta}} \cos(\theta/2) &  -i e^{i(\tilde{\omega}_\text{res} t_{\theta}+\phi)}\sin(\theta/2)\\
- i e^{-i \phi} \sin(\theta/2) &  \cos(\theta/2)
\end{pmatrix},
\end{equation}
where $t_{\theta} = \theta/\Omega$ is the duration of the rotation.
The leading order contribution to the fast oscillations which we treat here as infidelity, occurred at $O[\gamma^1]$, and are given by
\begin{equation}
\label{Eq:U01}
U_0^{(1)} = 2 i
\begin{pmatrix}
\cos(\tilde{\omega}_\text{res} t_{\theta}+2\phi)\sin(\theta/2)  &  e^{-i \phi}\sin(\tilde{\omega}_\text{res} t_{\theta}) \cos(\theta/2)\\
 e^{i(\tilde{\omega}_\text{res} t_{\theta}+\phi)} \sin(\tilde{\omega}_\text{res} t_{\theta}) \cos(\theta/2) &  -e^{i\tilde{\omega}_\text{res} t_{\theta}} \cos(\tilde{\omega}_\text{res} t_{\theta}+2\phi) \sin(\theta/2)
\end{pmatrix}.
\end{equation}

We now calculate the $\chi$ matrix of $\mathcal{E}_\text{sys}$ order by order, $\chi_\text{sys} = \chi_\text{ideal} + \sum_{n=1}^{\infty}\gamma^n \chi_\text{sys}^{(n)}$.
Since $\mathcal{E}_\text{sys}$ describes the unitary evolution (without decoherence), we must have $1 = \text{Tr}(\chi_\text{sys}\chi_\text{sys}) = 1 + 2 \gamma \text{Tr}(\chi_\text{ideal}\chi_\text{sys}^{(1)}) + \gamma^2 [\text{Tr}(\chi_\text{sys}^{(1)}\chi_\text{sys}^{(1)})+ 2 \, \text{Tr}(\chi_\text{ideal}\chi_\text{sys}^{(2)})] + \cdots$.
Equating terms order by order in $\gamma$, we arrive at
\begin{eqnarray}
&& \label{Eq:FirstOrder}
\text{Tr}(\chi_\text{ideal}\chi_\text{sys}^{(1)}) = 0, \\
&& \label{Eq:SecondOrder}
\text{Tr}(\chi_\text{ideal}\chi_\text{sys}^{(2)}) = -1/2 \, \text{Tr}(\chi_\text{sys}^{(1)}\chi_\text{sys}^{(1)}), \\
&& \label{Eq:ThirdOrder}
\text{Tr}(\chi_\text{ideal}\chi_\text{sys}^{(3)}) = -\, \text{Tr}(\chi_\text{sys}^{(1)}\chi_\text{sys}^{(2)}), \\
&& \label{Eq:FourthOrder}
\text{Tr}(\chi_\text{ideal}\chi_\text{sys}^{(4)}) = -\, \text{Tr}(\chi_\text{sys}^{(1)}\chi_\text{sys}^{(3)}) -1/2\, \text{Tr}(\chi_\text{sys}^{(2)}\chi_\text{sys}^{(2)}).
\end{eqnarray}
Using Eq.~(\ref{Eq:ProcessFidelity}) and Eq.~(\ref{Eq:FirstOrder})-(\ref{Eq:FourthOrder}) and applying our analytical results for $U_0$, i.e. Eqs.~(\ref{Eq:U00}) and (\ref{Eq:U01}) and high-order contributions, the fidelity $F$ up to $O[\gamma^3]$ is given by
\begin{equation}
\label{Eq:FidelityAt}
F_{\theta}(\phi) = 1 - 2 \gamma ^2 [1 - \cos ^2(\theta /2)\cos(2 \theta \tilde{\omega}_\text{res}/\Omega ) + \sin ^2(\theta /2)\cos(2 \theta \tilde{\omega}_\text{res}/\Omega + 4\phi)] -4 \gamma^3\sin(\theta)\sin(2 \theta \tilde{\omega}_\text{res}/\Omega).
\end{equation} 
If $\phi = (2m+1)\pi/4$ for any integer $m$, this expression can be simplified to yield Eq.~(2) in the main text:
\begin{equation}
\label{Eq:Fidelity3}
F_{\theta}(\phi) = 1 - 2 \gamma ^2 [1 - \cos(N \pi)] - 4 \gamma^3\sin(\theta)\sin(N \pi),
\end{equation}
where $N \equiv (2 \theta \tilde{\omega}_\text{res})/(\pi \Omega)$.
Note here that $F_{\theta}(\phi)$  vanishes up to $O[\gamma^3]$ whenever $N$ is an even integer.
The $O[\gamma^4]$ contribution for $\phi = (2m+1)\pi/4$ for any integer $m$ is
\begin{eqnarray}
\label{Eq:Fidelity4}
F_{\theta}^{(4)}(\pi/4) &=& 16[1-\cos(2 \theta \tilde{\omega}_\text{res}/\Omega )] -16\sin^2(\theta/2)\cos(4 \theta \tilde{\omega}_\text{res}/\Omega )+2 (-1)^{m} \cos^2(\theta/2)\sin(4 \theta \tilde{\omega}_\text{res}/\Omega ) \nonumber \\
&& -(5/2) [1-\cos(8 \theta \tilde{\omega}_\text{res}/\Omega )] -(-1)^{m} \sin (8 \theta \tilde{\omega}_\text{res}/\Omega ),
\end{eqnarray}
which can be simplified as $F_{\theta}^{(4)}(\phi) = -16 \sin^2(\theta/2)$ when $N$ is an even integer.
This represents an upper bound on the gate fidelity in any dip, as discussed in Sec.~\ref{section:RthetaGates} below, and in the main text.

We also note that, for arbitrary $\phi$ and any integer $k$, the fidelity can be written as
\begin{equation}
 F_{\theta}(\phi) = 1-
\left\{ 
\begin{array}{ll}
2 \gamma ^2 [1 - \cos(2 \theta \tilde{\omega}_\text{res}/\Omega )] + O[\gamma^4] & \text{if } \theta = 2 k \pi  \\
2 \gamma ^2 [1 - \cos(2 \theta \tilde{\omega}_\text{res}/\Omega + 4\phi)] + O[\gamma^4] & \text{if } \theta = (2 k +1) \pi 
\end{array} \right. ,
\end{equation} 
which equals unity, corresponding to perfect fidelity, whenever $(\theta \tilde{\omega}_\text{res}/\Omega)$ is an integer multiple of $\pi$ for $\theta = 2 k \pi$, or whenever $(\theta \tilde{\omega}_\text{res}/\Omega + 4\phi)$ is an integer multiple of $\pi$ for $\theta = (2 k +1) \pi$.

\section{Analytic Formula For Dynamics In The Presence Of Detuning Noise With Different Spectra}
\label{section:AsymptoticFormula}
In this section we analytically solve the dynamics of a strongly driven two-level system in the presence of classical detuning noise with different spectra. 
We first outline the derivation, and then present the results for noise with a generic spectrum.
We then discuss in detail the asymptotic formulas that describe the large-time behavior for three different spectra: Markovian noise, quasi-static noise, and $1/f$ noise.

The dynamics of a strongly driven two-level system coupled to classical noise affecting the detuning $\varepsilon$ can be described using the Hamiltonian
\begin{equation}
\label{Eq: HamiltonianTwoLevelSystem}
\Hamiltonian = \Hamiltonian_\text{sys} + \Hamiltonian_\text{n}  = - (\hbar \omega_\text{q}/2) \, \sigma_z + [A_\text{t} \sigma_x + A_{\ell} \sigma_z] \cos(\omega_\text{d} t + \phi) + [h_{\text{n},x} \sigma_x + h_{\text{n},z} \sigma_z] \delta \varepsilon(t),
\end{equation} 
where $\omega_\text{q}$ is the qubit angular frequency, $\omega_\text{d}$ is the angular frequency of the drive, $A_\text{t}$ ($A_{\ell}$) is the transverse (longitudinal) amplitude of the drive, and $h_{\text{n},x}$ ($h_{\text{n},z}$) is the transverse (longitudinal) coupling to the detuning noise $\delta \varepsilon$.
Here we start with a more general situation to include, taking a charge qubit as an example, the case of driving tunnel coupling at the sweet spot ($A_{\ell} \neq 0 $ when $\varepsilon = 0$), and also the case of working away from the sweet spot ($\varepsilon,\, A_\text{t}, \, A_\ell, \, h_{\text{n},x}, \, h_{\text{n},z} \neq 0$).
In the main text, we only consider driving detuning while working at the sweet spot, the special case where $\hbar \omega_\text{q} = 2 \Delta$, $A_{\ell} = 0$, $h_{\text{n},x} = -1/2$, and $h_{\text{n},z} = 0$.

In the absence of noise, the evolution operator $U_0$ satisfies the equation 
\begin{equation}
i \hbar \frac{d}{dt} U_0 = \Hamiltonian_\text{sys} U_0,
\end{equation}
where $U_0(t=0) = I$.
In Ref.~\cite{PhysRevA.95.062321}, we showed how to obtain $U_0$ order-by-order in the perturbation parameter $\gamma \sim A/\hbar\omega_\text{d} $, where $A = A_\text{t},\, A_{\ell}$, using a dressed-state formalism.
Transforming into the interaction picture, the equation of motion in the presence of the detuning noise can be written as 
\begin{equation}
i \hbar \frac{d}{dt} \rho^I = \delta \varepsilon(t) \mathcal{L} \rho^I ,
\end{equation}
where $\rho^I = U_0^{\dagger} \rho U_0$ is the density matrix in the interaction picture, and $\mathcal{L} \rho^I \equiv [h_\text{n}^I,\rho^I]$ with $h_\text{n}^I = U_0^{\dagger} h_\text{n} U_0$ and $h_\text{n} = h_{\text{n},x} \sigma_x + h_{\text{n},z} \sigma_z$.
The evolution can be expressed as a cumulant expansion~\cite{doi:10.1143/JPSJ.17.1100,doi:10.1063/1.1703941}
\begin{equation}
\langle \rho^I(t) \rangle = 
\exp \left\{ \sum\limits_{n=1}^{\infty} \frac{(-i)^n}{\hbar^n} \int\limits_0^t dt_1 \cdots \int\limits_0^{t_{n-1}} dt_n  \langle \mathcal{L}(t_1) \delta\varepsilon(t_1) \cdots \mathcal{L}(t_n) \delta\varepsilon(t_n) \rangle_c \right\} \rho^I(0)
,
\end{equation}
where $\langle \cdots \rangle$ is the ensemble average over $\delta\varepsilon(t)$ and $\langle \cdots \rangle_c$ is the cumulant average.
To $O[(\delta \varepsilon/\hbar \Omega)^2]$, this can be written as
\begin{equation}
\label{Eq:AnalyticFormula2nd}
\langle \rho^I(t) \rangle = 
\exp \left\{-\frac{1}{\hbar^2} \int\limits_0^t dt_1 \int\limits_0^{t_1} dt_2 \mathcal{L}(t_1)  \mathcal{L}(t_2) S(t_1-t_2)  \right\} \rho^I(0),
\end{equation}
where $\langle \cdots \rangle$ describes the ensemble average over $\delta\varepsilon(t)$, and $S(t_1-t_2) \equiv \langle \delta\varepsilon(t_1) \delta\varepsilon(t_2) \rangle$ is the time autocorrelation function of $\delta \varepsilon(t)$. We note that the power spectrum density of the noise, $\tilde S(\omega)$, is related to $S(t)$ by~\cite{RevModPhys.82.1155}
\begin{equation}
\label{Eq:Sw}
\tilde{S}(\omega) = \int\limits_{-\infty}^{\infty} dt\,e^{i \omega t} S(t).
\end{equation}

To simplify the calculation, we express the qubit state as a Bloch vector $\mathbf{r}^I$ in the interaction frame, [$\rho^I = 1/2 (I_2 +\mathbf{r}^I_x \sigma_x + \mathbf{r}^I_y \sigma_y + \mathbf{r}^I_z \sigma_z)$], and the super-operator $\mathcal{L}$ as a matrix by using $\{\sigma_x, \sigma_y, \sigma_z\}$ as the basis of its domain (two-by-two Hermitian operator with vanishing trace).  
Eq.~(\ref{Eq:AnalyticFormula2nd}) can then be written as
\begin{equation}
\label{Eq:Rt}
\mathbf{r}^I(t) = \exp [K(t) ] \mathbf{r}^I(0).
\end{equation}
By expressing $h_\text{n}^I(t)$ in terms of the Pauli matrices, $h_\text{n}^I(t) = h_{\text{n},x}^I(t) \sigma_x + h_{\text{n},y}^I(t) \sigma_y + h_{\text{n},z}^I(t) \sigma_z$, and further expanding the coefficients in Fourier series, $h_{\text{n},i}^I(t) =\sum\limits_{i,\omega}\alpha_{i,\omega} e^{i\omega t}$, the matrix $K(t)$ in Eq.~(\ref{Eq:Rt}) can be written as
\begin{eqnarray}
\label{Eq:R(t)}
[K(t)]_{ij} &=& -\frac{4}{\hbar^2} \int\limits_0^t dt_1 \int\limits_0^{t_1} dt_2 [\red{-} h_{\text{n},j}^I(t_1)h_{\text{n},i}^I(t_2)  \red{+} \mathbf{h}_\text{n}^I(t_1) \cdot \mathbf{h}_\text{n}^I(t_2) \delta_{ij} ] S(t_1-t_2)  \nonumber \\
&=& \frac{4}{\hbar^2} \sum\limits_{\omega_1, \omega_2}
[\red{-} \alpha_{j,\omega_1}\alpha_{i,\omega_2}  \red{+} \delta_{ij} \sum\limits_{k=1}^3\alpha_{k,\omega_1}\alpha_{k,\omega_2}] I(t,\omega_1,\omega_2), 
\end{eqnarray}
where
\begin{eqnarray}
\label{Eq:Integral}
I(t,\omega_1,\omega_2) &\equiv & \int\limits_0^t dt_1 \int\limits_0^{t_1} dt_2 e^{i\omega_1 t_1} e^{i\omega_2 t_2} S(t_1-t_2) , \\
\mathbf{h}_\text{n}^I &\equiv & [h_{\text{n},x}^I,h_{\text{n},y}^I,h_{\text{n},z}^I].
\end{eqnarray}
In general, $K(t) = K_\text{M} + K_{\varphi} + K_{\text{nMn}\varphi}$ where $K_\text{M}$ is a Markovian term and $K_{\varphi}$ is a pure-dephasing term~\cite{PhysRevLett.88.228304,PhysRevLett.93.267007, PhysRevB.72.134519}. We denote the remaining term, $K_{\text{nMn}\varphi}$, as the non-Markovian-non-dephasing term; it is the correction to the Markovian approximation. 
The Markovian term is defined as the decoherence within the Markovian approximation (i.e., when the correlation time is much smaller than the characteristic time scale of the system dynamics). If only the Markovian term is present, the decoherence yields an exponential decay and $K(t)$ is linear in time $t$. For further discussion, please see Sec.~\ref{Sec:MarkovianNoise} and the main text.
The pure-dephasing term describes pure dephasing in the rotating frame defined by $U_\text{rot}$=$\text{diag}[e^{i\omega_\text{d} t/2},e^{-i\omega_\text{d} t/2}]$, and is associated with $I(t,\omega_1 = 0,\omega_2 = 0)$~\cite{PhysRevLett.88.228304,PhysRevLett.93.267007, PhysRevB.72.134519}.
The non-Markovian-non-dephasing term is the correction to the Markovian approximation, describing, for example, the decoherence due to low-frequency noise, and oscillatory terms typically ignored in the asymptotic limit $t \gg 2 \pi/\omega$, where $\omega$ represents any relevant angular frequency in the system.

In the following subsections, we evaluate the three terms of $K(t)$, $\{ K_\text{M},K_{\varphi},K_{\text{nMn}\varphi}\}$, for general $S(t_1-t_2)$, and then describe the asymptotic form of $K(t)$ for three cases: Markovian noise, quasistatic noise, and $1/f$ noise.
To be concise, we assume the qubit is driven resonantly ($\omega_\text{d} = \tilde{\omega}_\text{res}$).

We note that this technique is easy to use and is applicable to many other kinds of noise that are present in condensed matter devices such as Lorentzian noise and power-law noise, and can also be generalized to multi-level systems such as quantum-dot hybrid qubits~\cite{Shi2012,Koh2012,KimShiSimmonsEtAl2014,Ferraro2014,Cao2016,2016arXiv161104945T,PhysRevA.95.062321}. 

\subsection{Analytic formula for general noise spectra}
For a generic noise spectrum $\tilde{S}(\omega)$, we define the symmetric and antisymmetric versions of Eq.~(\ref{Eq:Integral}) as follows:
\begin{eqnarray}
I_S(t,\omega_1,\omega_2) 
&=& \frac{1}{2}\int_0^t dt_1 \int_0^{t_1} dt_2 \,  [e^{i\omega_1 t_1} e^{i\omega_2 t_2} + e^{i\omega_2 t_1} e^{i\omega_1 t_2}] S(t_1-t_2),  \\
I_A(t,\omega_1,\omega_2) 
&=& \frac{1}{2}\int_0^t dt_1 \int_0^{t_1} dt_2 \,  [e^{i\omega_1 t_1} e^{i\omega_2 t_2} - e^{i\omega_2 t_1} e^{i\omega_1 t_2}] S(t_1-t_2),  
\end{eqnarray}
such that $I(t,\omega_1,\omega_2) = I_\text{S}(t,\omega_1,\omega_2) + I_\text{A}(t,\omega_1,\omega_2)$.
The partially evaluated integrals for different cases of $\{\omega_1, \omega_2\}$ are given by
\begin{eqnarray}
\label{Eq:IS}
&& I_\text{S}(t,\omega_1,\omega_2)  \\ &&= \left\{ 
\begin{array}{ll}
- \left[f_\text{t}(t) - t f_\text{0}(t)\right] & (\omega_1 = \omega_2 = 0) ,  \\
-\left[f_\text{tcos}(t,\omega_1) - t f_\text{cos}(t,\omega_1)\right] & (\omega_1 = -\omega_2 \neq 0) ,   \\
-\frac{e^{i \frac{\omega_1}{2}t}}{\omega_1}\left[\cos(\frac{\omega_1}{2}t) f_\text{sin}(t,\omega_1) -\sin(\frac{\omega_1}{2}t) (f_\text{cos}(t,\omega_1)+f_\text{0}(t))	\right] & (\omega_1 \neq 0 , \, \omega_2 = 0) ,   \\
-\frac{e^{i \frac{ \omega_2}{2}t}}{ \omega_2}\left[\cos(\frac{ \omega_2}{2}t) f_\text{sin}(t,\omega_2) -\sin(\frac{ \omega_2}{2}t) (f_\text{0}(t) + f_\text{cos}(t,\omega_2))\right] & (\omega_1 = 0 , \, \omega_2\neq 0) ,  \\
 -\frac{e^{i \frac{\omega_1 + \omega_2}{2}t}}{\omega_1 + \omega_2} \left[\cos(\frac{\omega_1 + \omega_2}{2}t) (f_\text{sin}(t,\omega_1) + f_\text{sin}(t,\omega_2)) -\sin(\frac{\omega_1 + \omega_2}{2}t) (f_\text{cos}(t,\omega_1) + f_\text{cos}(t,\omega_2)\right] & (\text{Otherwise}) ,
\end{array}
\right.  \nonumber \\
\label{Eq:IA}
&&I_\text{A}(t,\omega_1,\omega_2)  \\ &&= \left\{ 
\begin{array}{ll}
0 & (\omega_1 = \omega_2 = 0) , \\
- i \left[f_\text{tsin}(t,\omega_1) - t f_\text{sin}(t,\omega_1)\right] & (\omega_1 = -\omega_2 \neq 0) , \\
-i\frac{e^{i \frac{\omega_1}{2}t}}{\omega_1} \left[-\cos(\frac{\omega_1}{2}t) (f_\text{cos}(t,\omega_1) - f_\text{0}(t)) -\sin(\frac{\omega_1}{2}t) f_\text{sin}(t,\omega_1)	\right] & (\omega_1 \neq 0 , \,\omega_2 = 0) , \\
-i\frac{e^{i \frac{\omega_2}{2}t}}{\omega_2} \left[-\cos(\frac{\omega_2}{2}t) (f_\text{0}(t) - f_\text{cos}(t,\omega_2) )+\sin(\frac{\omega_2}{2}t) f_\text{sin}(t,\omega_2)\right] & (\omega_1 = 0 , \, \omega_2\neq 0) , \\
i\frac{e^{i \frac{\omega_1 + \omega_2}{2}t}}{\omega_1 + \omega_2} \left[\cos(\frac{\omega_1 + \omega_2}{2}t) (f_\text{cos}(t,\omega_1)- f_\text{cos}(t,\omega_2))+\sin(\frac{\omega_1 + \omega_2}{2}t) (f_\text{sin}(t,\omega_1)-f_\text{sin}(t,\omega_2))	\right] & (\text{Otherwise}) ,
\end{array}
\right. \nonumber
\end{eqnarray}
where
\begin{eqnarray}
\label{Eq:ftcos}
f_\text{tcos}(t,\omega) &=&  \int_0^t dt^\prime S(t^\prime) \, t^\prime \cos(\omega t^\prime), \\
f_\text{tsin}(t,\omega) &=& \int_0^t dt^\prime S(t^\prime) \, t^\prime \sin(\omega t^\prime), \\
f_\text{t}(t) & =& \int_0^t dt^\prime S(t^\prime) \, t^\prime, \\
f_\text{cos}(t,\omega) &=&  \int_0^t dt^\prime S(t^\prime) \, \cos(\omega t^\prime), \\
f_\text{sin}(t,\omega) &=& \int_0^t dt^\prime S(t^\prime) \, \sin(\omega t^\prime), \\
\label{Eq:f0}
f_\text{0}(t) &=& \int_0^t dt^\prime S(t^\prime). 
\end{eqnarray}
As long as one can evaluate Eq.~(\ref{Eq:ftcos})-(\ref{Eq:f0}), or approximate these equations to within the desired accuracy, one can obtain an analytic formula for the dynamics in the interaction frame using Eq.~(\ref{Eq:IS}), Eq.~(\ref{Eq:IA}), and Eq.~(\ref{Eq:R(t)}).
This formula can be applied to analytically describe the system dynamics, including decoherence, or to accurately estimate the quantum gate fidelity.
For example, Figs.~1(b) and 1(c) in the main text compare the analytic formula and simulation of the dynamics of a strongly driven charge qubit coupled to $1/f$ charge noise.
The analytic formula captures the main features of the simulation; the deviation of the two is smaller than $10^{-3}$ over the time shown in the figure.
Figs.~3(a) and 3(b) in the main text compare the analytical results and the simulation results of the gate fidelity $F_{\pi}(\phi = \pi/4)$. 
The analytical results match the simulation results in the regimes of greatest interest, for example when the driving amplitude is large.
In Fig.~3(a), deviations between theory and simulations begin to appear when the driving amplitude is small because the higher order effects of the noise become visible when the gate time is long. 
In Fig.~3(b), deviations also arise from the higher order terms in the noise $\delta \varepsilon$ when the tunnel coupling is small. 
When the tunnel coupling is large, deviation appears again because of the higher order terms in $\gamma$. 
However, we note that the deviations are quite small, on the order of $10^{-4}$, and are mainly visible here because the data are plotted on a log scale. 

In order to describe the asymptotic behavior of the dynamics at long times, we calculate the asymptotic form of $K(t)$ by taking the limit $t \gg 2 \pi/\omega$, where $\omega$ represents any relevant angular frequency in the system, and neglecting the oscillatory terms. 
The asymptotic forms of $K_\text{M}(t)$, $K_{\varphi}(t)$, and $K_{\text{nMn}\varphi}(t)$ typically can be expressed as  
\begin{eqnarray}
\label{Eq:RM}
[K_\text{M}]_{ij} &=& -\frac{4}{\hbar^2} \sum\limits_{\omega_1 }
[\red{-}\alpha_{j,\omega_1}\alpha_{i,-\omega_1} 
\red{+}\delta_{ij} \sum\limits_{k=1}^3\alpha_{k,\omega_1}\alpha_{k,-\omega_1} ] I_S(t,\omega_1,-\omega_1), \\
\label{Eq:Rphi}
[K_{\varphi}]_{ij} &=& -\frac{4}{\hbar^2} 
[\red{-}\alpha_{j,0}\alpha_{i,0} 
\red{+}\delta_{ij} \sum\limits_{k=1}^3\alpha_{k,0}\alpha_{k,0} ] I_S(t,0,0), \\
\label{Eq:RnM}
[K_{\text{nMn}\varphi}]_{ij} &=& -\frac{4}{\hbar^2} \sum\limits_{\omega_1 }
[\red{-}\alpha_{j,\omega_1}\alpha_{i,-\omega_1} 
\red{+}\delta_{ij} \sum\limits_{k=1}^3\alpha_{k,\omega_1}\alpha_{k,-\omega_1} ] I_A(t,\omega_1,-\omega_1) .
\end{eqnarray}
We note that in some cases, $I(t,\omega_1 = 0,\omega_2\neq 0)$ also contributes to the asymptotic form of $K_{\text{nMn}\varphi}$.

In the following, we focus on three typical types of noise: Markovian noise, quasi-static noise, and $1/f$ noise. We provide the results for the functions defined in Eqs.~(\ref{Eq:ftcos})-(\ref{Eq:f0}), the corresponding integrals $I(t,\omega_1,\omega_2)$, and the asymptotic formulas for these quantities. 
We express $K(t)$ as a perturbation series in $\gamma$, using $K_{x}^{(i)}$ to label the $O[\gamma^i]$ contribution to $K_{x}$, where $x$ can be M, $\varphi$, or nMn$\varphi$.

\subsection{Markovian Noise \label{Sec:MarkovianNoise}}
In the Markovian limit, we assume that the correlation time $\tau_c$ is much smaller than the time scale we are interested in, $t$, and $1/\omega$, where $\omega$ is the characteristic frequency of the system (i.e., for a resonantly driven charge qubit, the resonant angular frequency $\omega_\text{res}$, the Rabi angular frequency $\Omega$ or their linear combinations), such that $ t/(2\pi) \gg 1/\omega \gg \tau_c/(2 \pi)$. 
By setting $S(t^{\prime}) \approx 0$ when $t^{\prime} > \tau_c$ and neglecting any term that scales with $\tau_c$, Eqs.~(\ref{Eq:ftcos})-(\ref{Eq:f0}) can be evaluated to yield  
\begin{align}
\label{Eq:Markovianf1}
f_\text{tcos}(\omega) &\approx 0, \\
f_\text{tsin}(\omega) &\approx 0, \\
f_\text{t}(\omega) &\approx 0, \\
f_\text{cos}(\omega) &\approx \frac{1}{2} \tilde{S}(\omega), \\
f_\text{sin}(\omega) &\approx 0, \\
\label{Eq:Markovianf6}
f_\text{0}(\omega) &\approx \frac{1}{2} \tilde{S}(0).
\end{align}
Inserting Eq.~(\ref{Eq:Markovianf1}) to Eq.~(\ref{Eq:Markovianf6}) in Eq.~(\ref{Eq:IS}) and Eq.~(\ref{Eq:IA}) and neglecting $O[1/\omega]$ terms, we obtain $I = I_\text{S} + I_\text{A}$, with:
\begin{eqnarray}
\label{Eq:ISMarkovian}
I_\text{S}(t,\omega_1,\omega_2) &\approx & \left\{
\begin{array}{ll}
\tilde{S}(\omega_1) t/2  & (\omega_1 = -\omega_2) , \\
0 & (\text{Otherwise}) ,
\end{array}
\right.  \\
\label{Eq:IAMarkovian}
I_\text{A}(t,\omega_1,\omega_2) &\approx & 0 .
\end{eqnarray}

Applying this formula to Eq.~(\ref{Eq: HamiltonianTwoLevelSystem}) and only keeping the terms up to $O[\gamma^0]$ (consistent with the rotating wave approximation) yields
\begin{equation}
K^{(0)}(t) = K_\text{M}^{(0)}(t) =
-t
\begin{pmatrix}
\Gamma_{x} & \Gamma_{xy} & 0 \\
\Gamma_{xy} & \Gamma_{y} & 0 \\
0 & 0 & \Gamma_{z}
\end{pmatrix}   , \\ 
\end{equation}
where 
\begin{eqnarray}
&&\Gamma_{x} = \frac{h_{\text{n},z}^2}{\hbar^2} \frac{3+\cos(2\phi)}{2}\tilde{S}(\Omega) +  \frac{h_{\text{n},x}^2}{\hbar^2} [ \sin^2(\phi) \tilde{S}(\tilde{\omega}_\text{res})+\frac{1+\cos^2(\phi)}{4}\tilde{S}(\tilde{\omega}_\text{res}+\Omega)+ \frac{1+\cos^2(\phi)}{4} \tilde{S}(\tilde{\omega}_\text{res}-\Omega)] , \\
&&\Gamma_{y} = 
\frac{h_{\text{n},z}^2}{\hbar^2}\frac{3-\cos(2\phi)}{2} \tilde{S}(\Omega) + \frac{h_{\text{n},x}^2}{\hbar^2}[\cos^2(\phi) \tilde{S}(\tilde{\omega}_\text{res}) +  \frac{2-\cos^2(\phi)}{4} \tilde{S}(\tilde{\omega}_\text{res}+\Omega) + \frac{2-\cos^2(\phi)}{4} \tilde{S}(\tilde{\omega}_\text{res}-\Omega)]  ,\\
&&\Gamma_{z} = 
\frac{h_{\text{n},z}^2}{\hbar^2} \tilde{S}(\Omega) + \frac{h_{\text{n},x}^2}{\hbar^2}[\tilde{S}(\tilde{\omega}_\text{res}) +  \frac{1}{4}\tilde{S}(\tilde{\omega}_\text{res}+\Omega)+ \frac{1}{4} \tilde{S}(\tilde{\omega}_\text{res}-\Omega)]  ,\\
&&\Gamma_{xy} = -\frac{h_{\text{n},z}^2}{2\hbar^2} \sin(2\phi) \tilde{S}(\Omega) + \frac{h_{\text{n},x}^2}{2\hbar^2} \sin(2\phi) [\tilde{S}(\tilde{\omega}_\text{res}) -  \frac{1}{4}\tilde{S}(\tilde{\omega}_\text{res}+\Omega) - \frac{1}{4} \tilde{S}(\tilde{\omega}_\text{res}-\Omega)], 
\end{eqnarray}
and $K_{\varphi} = K_{\text{nMn}\varphi} = 0$.
Here, $\Omega$ is the Rabi angular frequency.
If we apply this result on the detuning-driven charge qubit at the sweet spot ($\varepsilon = 0$), as discussed in the main text, and focus on the small drive regime, $\tilde \omega_\text{res} \gg \Omega$, the results can be further simplified as
 \begin{equation*}
\Gamma_{x} = \frac{h_{\text{n},x}^2}{\hbar^2} [ 1+\frac{1}{2}\sin^2(\phi) ]\tilde{S}(\tilde{\omega}_\text{res}),\,
\Gamma_{y} = \frac{h_{\text{n},x}^2}{\hbar^2} [ 1+\frac{1}{2}\cos^2(\phi) ]\tilde{S}(\tilde{\omega}_\text{res}),\, 
\Gamma_{z} = \frac{3h_{\text{n},x}^2}{2\hbar^2}\tilde{S}(\tilde{\omega}_\text{res}), \,
\Gamma_{xy} =  \frac{h_{\text{n},x}^2}{4\hbar^2} \sin(2\phi) \tilde{S}(\tilde{\omega}_\text{res}).
 \end{equation*}

Note that, if $\phi = 0$, the expressions for $\Gamma_{x}$ and $\Gamma_{z}=\Gamma_{y}$ are equivalent to those obtained using Bloch-Redfield theory~\cite{PhysRevB.72.134519,RevModPhys.82.1155, Slitcher1990}.
For this special case, the elements of the Bloch vector exhibit an exponential decay,
\begin{equation}
\mathbf{r}^I(t) = 
\begin{pmatrix}
e^{-\Gamma_{x} t} & 0 & 0 \\
0 & e^{-\Gamma_{z} t} & 0 \\
0 & 0 & e^{-\Gamma_{z} t}
\end{pmatrix}
\mathbf{r}^I(0) .
\end{equation}

If one now restore the $O(1/\Omega)$ terms, previously neglected while deriving Eqs.~(\ref{Eq:ISMarkovian}) and (\ref{Eq:IAMarkovian}), one obtains an equation of motion for the Bloch vector within the rotating wave approximation,
\begin{equation}
\label{Eq:MasterEquation}
\frac{d}{dt}\mathbf{r}^I(t) = 
\begin{pmatrix}
 -\Gamma_{x}  & 0 & 0 \\
 \eta\sin(\Omega\, t) & -\Gamma_{z}  - \lambda \cos(2\Omega\, t)  &  \sgn(A_\text{t})\, \lambda \sin(2\Omega\, t)  \\
 \sgn(A_\text{t})\,\eta \cos(\Omega\, t) &  \sgn(A_\text{t})\, \lambda \sin(2\Omega\, t) & -\Gamma_{z}  + \lambda \cos(2\Omega\, t)
\end{pmatrix}\mathbf{r}^I(t) ,
\end{equation}
where $\eta = \frac{1}{2} h_{n,x}^2[\tilde{S}(\tilde{\omega}_\text{res}+\Omega)-\tilde{S}(\tilde{\omega}_\text{res}-\Omega)]$, $\lambda = -\frac{1}{4}h_{n,x}^2[\tilde{S}(\tilde{\omega}_\text{res}+\Omega)+\tilde{S}(\tilde{\omega}_\text{res}-\Omega)] + h_{n,z}^2 \tilde{S}(\Omega)$, and $\sgn(x) = +1$ ($-1$) when $x>0$ ($x<0$).
We note that Equation~(\ref{Eq:MasterEquation}) is equivalent to the master equation results derived in~\cite{doi:10.1063/1.468844}.
However, the current method also provides a systematic approach for including higher-order terms in $\gamma$, beyond the rotating-wave approximation.

\subsection{Quasistatic Noise}
For quasistatic noise with standard deviation $\sigma_{\varepsilon}$, the time correlation function is  
\begin{equation}
S(t) = \sigma_{\varepsilon}^2,
\end{equation}
and the corresponding noise spectral density is
\begin{equation}
\label{Eq:Spectrum Quasi-static1}
\tilde{S}(\omega) = 2 \pi \sigma_{\varepsilon}^2 \delta(\omega).
\end{equation}
Using these forms in Eqs.~(\ref{Eq:ftcos})-(\ref{Eq:f0}) yields  
\begin{align}
f_\text{tcos}(t,\omega) & = \sigma_{\varepsilon}^2 \frac{-1 + \cos (\omega t) + \omega t \sin(\omega t)}{\omega^2}, \\
f_\text{tsin}(t,\omega) & = \sigma_{\varepsilon}^2 \frac{-\omega t \cos (\omega t) + \sin(\omega t)}{\omega^2}, \\
f_\text{t}(t) & = \sigma_{\varepsilon}^2 \frac{t^2}{2}, \\
f_\text{cos}(t,\omega) & =  \sigma_{\varepsilon}^2 \frac{\sin(\omega t)}{\omega}, \\
f_\text{sin}(t,\omega) & =  \sigma_{\varepsilon}^2 \frac{1-\cos(\omega t)}{\omega}, \\
f_\text{0}(t) & =\sigma_{\varepsilon}^2  t.
\end{align}

Taking the limit $t/(2\pi) \gg 1/\omega$ for any relevant angular frequencies of the system $\omega$, and keeping only the non-oscillatory terms to $O[1/\omega]$, the integrals can be approximated as 
\begin{equation}
I(t,\omega_1,\omega_2) \approx \left\{
\begin{array}{ll}
\sigma_{\varepsilon}^2 t^2/2 & (\omega_1 = \omega_2 = 0) , \\
i\, \sigma_{\varepsilon}^2 \, t/\omega_1 & (\omega_1 = -\omega_2 \neq 0) , \\
i\, \sigma_{\varepsilon}^2 \, t/\omega_2 & (\omega_1 = 0 \text{ and } \omega_2 \neq 0) , \\
0 & (\text{Otherwise}) .
\end{array}
\right. \\
\end{equation}

For quasistatic noise, $K(t) =  K_{\varphi} +K_{\text{nMn}\varphi}(t)$, and $K_\text{M} = 0$.
The leading order term in $K_{\text{nMn}\varphi}(t)$ is $O[\gamma^0]$: 
\begin{equation}
K_{\text{nMn}\varphi}^{(0)}(t) =
-\Gamma_{\text{nMn}\varphi} t
\begin{pmatrix}
0 & 0 & \sin(\phi) \\
0 & 0 & \cos(\phi) \\
-\sin(\phi) & -\cos(\phi) & 0
\end{pmatrix}  ,
\end{equation}
where 
\begin{eqnarray}
&&\Gamma_{\text{nMn}\varphi} = -\frac{i}{2 \hbar^2} \sgn(A_\text{t}) [4 h_{\text{n},z}^2 \tilde{S}_\text{imag}(\Omega)+  
h_{\text{n},x}^2(\tilde{S}_\text{imag}(-\tilde{\omega}_\text{res} + \Omega)+\tilde{S}_\text{imag}(\tilde{\omega}_\text{res} + \Omega))]  ,\\
&& \tilde{S}_\text{imag}(\omega) \equiv  \frac{i\, \sigma_{\varepsilon}^2}{\omega}  . 
\end{eqnarray}
The leading order contribution to $K_{\varphi}(t)$ is $O[\gamma^2]$: 
\begin{equation}
 K_{\varphi}^{(2)}(t) = 
-\Gamma_{\varphi} t^2 
\begin{pmatrix}
\sin^2(\phi) & \cos(\phi)\sin(\phi) & 0 \\
\cos(\phi)\sin(\phi) & \cos^2(\phi) & 0 \\
0 & 0 & 1
\end{pmatrix} ,
\end{equation}
where 
\begin{equation}
\Gamma_{\varphi} = 2 \sigma_{\varepsilon}^2 h_{\text{n}x}^2 \frac{A_{\ell}^2}{\hbar^2\tilde{\omega}_\text{res}^2}.
\end{equation}
We note that, for the charge qubit operated at the sweet spot as discussed in the main text, this term is non-vanishing only if the tunnel coupling driving amplitude is not zero. 
This indicates that one should only drive detuning in order to suppress this decoherence effect, as in the main text.

To provide a physical picture, we note that $\hbar d \Omega / d\varepsilon = 2 h_{\text{n}x} A_{\ell} / \hbar \tilde{\omega}_\text{res}$.
By writing $\Gamma_{\varphi} = \sigma_{\varepsilon}^2 (\hbar d \Omega / d\varepsilon)^2/2 $, we interpret $K_{\varphi}(t)$ as the dephasing in the rotating frame, where the eigenenergies of the dressed states are $\pm \hbar \Omega$~\cite{PhysRevB.72.134519}.

These two terms cause the elements of the Bloch vector to decay as a Gaussian with a sinusoidal oscillation:
\begin{equation}
\mathbf{r}^I(t) = 
\begin{pmatrix}
1 + [e^{-\Gamma_{\varphi} t^2} \cos(\Gamma_{\text{nMn}\varphi} t)-1]\sin^2(\phi) & [e^{-\Gamma_{\varphi} t^2} \cos(\Gamma_{\text{nMn}\varphi} t)-1]\cos(\phi)\sin(\phi) & - e^{-\Gamma_{\varphi} t^2} \sin(\Gamma_{\text{nMn}\varphi} t)\sin(\phi)\\
[e^{-\Gamma_{\varphi} t^2} \cos(\Gamma_{\text{nMn}\varphi} t)-1]\cos(\phi)\sin(\phi) & 1 + [e^{-\Gamma_{\varphi} t^2} \cos(\Gamma_{\text{nMn}\varphi} t)-1]\cos^2(\phi) &  - e^{-\Gamma_{\varphi} t^2} \sin(\Gamma_{\text{nMn}\varphi} t)\cos(\phi)\\
e^{-\Gamma_{\varphi} t^2} \sin(\Gamma_{\text{nMn}\varphi} t)\sin(\phi) & e^{-\Gamma_{\varphi} t^2} \sin(\Gamma_{\text{nMn}\varphi} t)\cos(\phi) &  e^{-\Gamma_{\varphi} t^2} \cos(\Gamma_{\text{nMn}\varphi} t)
\end{pmatrix}
\mathbf{r}^I(0) .
\end{equation}


\subsection{$1/f$ noise \label{Sec:OneOverF}}
The $1/f$ power spectrum was presented in Eq.~(3) of the main text
\begin{equation}
\label{Eq:Spectrum One-Over-f}
\tilde{S}(\omega) = \left\{
  \begin{array}{cl}
    c_{\varepsilon}^2 \frac{2 \pi}{|\omega|} & : \omega_l \leq |\omega | \leq \omega_h\\
    0 & : \text{otherwise} 
  \end{array}
\right.,
\end{equation}
where $\omega_l$ ($\omega_h$) is the low (high) angular frequency cut-off.
The corresponding time correlation function is
\begin{equation}
S(t) = 2 c_{\varepsilon}^2 \left[ \Ci (\omega_h |t|) - \Ci (\omega_l |t|) \right],
\end{equation}
where $\Ci(x) \equiv -\int_x^{\infty} dx^{\prime} \cos(x^{\prime})/x^{\prime}$ is the cosine integral function, and $\Si(x) \equiv \int_0^{x} dx^{\prime} \sin(x^{\prime})/x^{\prime}$ is the sine integral function.
We further define functions
\begin{align}
\Ciw(\omega) &\equiv 2 \left[ \Ci \left(|(\omega+\omega_h)t|\right) + \Ci \left(|(\omega-\omega_h)t|\right) - \Ci \left(|(\omega+\omega_l)t|\right) - \Ci \left(|(\omega-\omega_l)t|\right) \right], \\
\Siw(\omega) &\equiv 2 \left[ \Si \left((\omega+\omega_h)t\right) + \Si \left((\omega-\omega_h)t\right) - \Si \left((\omega+\omega_l)t\right) - \Si \left((\omega-\omega_l)t\right) \right],
\end{align}
and obtain
\begin{align}
f_\text{tcos}(t,\omega) =\,& \frac{1}{\omega}t\sin(\omega t)S(t)
+ \frac{1}{\omega}\left\{ \frac{1}{\omega}\cos(\omega t)S(t) - \frac{c_{\varepsilon}^2}{2\omega}\left[\Ciw(\omega) - 2  \log\left(\left|\frac{\omega_h^2-\omega^2}{\omega_h^2} \frac{\omega_l^2}{\omega_l^2-\omega^2} \right|\right)\right]\right\} \nonumber \\
- \, &\frac{c_{\varepsilon}^2}{\omega}\left[
\frac{1-\cos ((\omega+\omega_h)t)}{\omega+\omega_h} + \frac{1-\cos ((\omega-\omega_h)t)}{\omega-\omega_h} - \frac{1-\cos ((\omega+\omega_l)t)}{\omega+\omega_l} -  \frac{1-\cos ((\omega-\omega_l)t)}{\omega-\omega_l}\right] 
, \\
f_\text{tsin}(t,\omega) =\,& -\frac{1}{\omega} t \cos(\omega t)S(t) + \frac{1}{\omega}\left\{\frac{1}{\omega}\sin(\omega t)S(t) - \frac{c_{\varepsilon}^2}{2\omega}\Siw(\omega) \right\} \nonumber \\
 \,&  +\frac{c_{\varepsilon}^2}{\omega}\left[
\frac{\sin((\omega+\omega_h)t)}{\omega+\omega_h} + \frac{\sin((\omega-\omega_h)t)}{\omega-\omega_h} - \frac{\sin((\omega+\omega_l)t)}{\omega+\omega_l} -  \frac{\sin((\omega-\omega_l)t)}{\omega-\omega_l}\right]
, \\
f_\text{t}(t) =\,& \frac{1}{2} t^2 S(t)- c_{\varepsilon}^2 t \left[\frac{\sin(\omega_h t)}{\omega_h} - \frac{\sin(\omega_l t)}{\omega_l}\right] + c_{\varepsilon}^2 \left[\frac{1-\cos (\omega_h t)}{\omega_h^2} -  \frac{1-\cos (\omega_l t)}{\omega_l}\right],\\
f_\text{cos}(t,\omega) =\,& \frac{1}{\omega}\sin(\omega t)S(t)- \frac{c_{\varepsilon}^2}{2\omega}\Siw(\omega) , \\
f_\text{sin}(t,\omega) =\,& -\frac{1}{\omega}\cos(\omega t)S(t) + \frac{c_{\varepsilon}^2}{2\omega}\left[\Ciw(\omega) - 2  \log\left(\left|\frac{\omega_h^2-\omega^2}{\omega_h^2} \frac{\omega_l^2}{\omega_l^2-\omega^2} \right|\right)\right] , \\
f_\text{0}(t) =\,&  t S(t) - 2 c_{\varepsilon}^2 \left[\frac{\sin(\omega_h t)}{\omega_h} - \frac{\sin(\omega_l t)}{\omega_l}\right].
\end{align}


We can obtain an asymptotic decay formula for $1/f$ noise by taking the long time limit, $1/\omega_l \gg t/(2\pi) \gg 1/\omega \gg 1/\omega_h$, where $t$ is the qubit evolution time we are interested in and $\omega$ is any angular frequency relevant to the system, and dropping the oscillatory terms in the full formula. 
The integrals $I(t,\omega_1,\omega_2)$ can be approximated as 
\begin{equation}
\label{Eq:IntergalOneOverF}
I(t,\omega_1,\omega_2) 
\approx \left\{
\begin{array}{ll}
c_{\varepsilon}^2 t^2 \left[\log(1/\omega_l t) -\gamma_E+ 3/2 \right] & (\omega_1 = \omega_2 = 0),  \\
c_{\varepsilon}^2 (\pi + 2 i \log|\omega_1/\omega_l|) t / \omega_1 & (\omega_1 = - \omega_2 \neq 0),  \\
2 i\, c_{\varepsilon}^2 t(\log(1/\omega_l t) -\gamma_E+ 1)/\omega_2& (\omega_1 = 0 \text{ and } \omega_2 \neq 0),  \\
0 & (\text{Otherwise}) ,
\end{array}
\right.  \nonumber
\end{equation}
where $\gamma_E\approx 0.577$ is the Euler's constant.

For a resonantly driven two-level system, only $K_\text{M}(t)$ and $K_{\text{nMn}\varphi}(t)$ have contributions of $O[(A/\hbar \omega_\text{d})^0]$, corresponding to the rotating-wave approximation.  
The lowest-order contributions to $K_{\varphi}(t)$ is $O[(A/\hbar \omega_\text{d})^2]$.
Below, we report the leading order contributions of $K_\text{M}(t)$, $K_{\text{nMn}\varphi}(t)$, and $K_{\varphi}(t)$, respectively.
The explicit expressions of $O[\gamma]$ contributions of $K_\text{M}(t)$, $K_{\text{nMn}\varphi}(t)$ are, however, omitted for brevity.

It is unsurprising that $1/f$ noise causes both Markovian and non-Markovian types of decoherence. 
At $O[\gamma^0]$, 
\begin{eqnarray}
K_\text{M}^{(0)}(t) &=&
-t
\begin{pmatrix}
\Gamma_{x} & \Gamma_{xy} & 0 \\
\Gamma_{xy} & \Gamma_{y} & 0 \\
0 & 0 & \Gamma_{z}
\end{pmatrix}   , \\ 
K_{\text{nMn}\varphi}^{(0)}(t) &=&
-\Gamma_{\text{nMn}\varphi} t
\begin{pmatrix}
0 & 0 & \sin(\phi) \\
0 & 0 & \cos(\phi) \\
-\sin(\phi) & -\cos(\phi) & 0
\end{pmatrix}  ,
\end{eqnarray}
where 
\begin{eqnarray}
&&\Gamma_{x} = \frac{h_{\text{n},z}^2}{\hbar^2} \frac{3+\cos(2\phi)}{2}\tilde{S}(\Omega) +  \frac{h_{\text{n},x}^2}{\hbar^2} [ \sin^2(\phi) \tilde{S}(\tilde{\omega}_\text{res})+\frac{1+\cos^2(\phi)}{4}\tilde{S}(\tilde{\omega}_\text{res}+\Omega)+ \frac{1+\cos^2(\phi)}{4} \tilde{S}(\tilde{\omega}_\text{res}-\Omega)] , \\
&&\Gamma_{y} = 
\frac{h_{\text{n},z}^2}{\hbar^2}\frac{3-\cos(2\phi)}{2} \tilde{S}(\Omega) + \frac{h_{\text{n},x}^2}{\hbar^2}[\cos^2(\phi) \tilde{S}(\tilde{\omega}_\text{res}) +  \frac{2-\cos^2(\phi)}{4} \tilde{S}(\tilde{\omega}_\text{res}+\Omega) + \frac{2-\cos^2(\phi)}{4} \tilde{S}(\tilde{\omega}_\text{res}-\Omega)]  ,\\
&&\Gamma_{z} = 
\frac{h_{\text{n},z}^2}{\hbar^2} \tilde{S}(\Omega) + \frac{h_{\text{n},x}^2}{\hbar^2}[\tilde{S}(\tilde{\omega}_\text{res}) +  \frac{1}{4}\tilde{S}(\tilde{\omega}_\text{res}+\Omega)+ \frac{1}{4} \tilde{S}(\tilde{\omega}_\text{res}-\Omega)]  ,\\
&&\Gamma_{xy} = -\frac{h_{\text{n},z}^2}{2\hbar^2} \sin(2\phi) \tilde{S}(\Omega) + \frac{h_{\text{n},x}^2}{2\hbar^2} \sin(2\phi) [\tilde{S}(\tilde{\omega}_\text{res}) -  \frac{1}{4}\tilde{S}(\tilde{\omega}_\text{res}+\Omega) - \frac{1}{4} \tilde{S}(\tilde{\omega}_\text{res}-\Omega)], \\
&&\Gamma_{\text{nMn}\varphi} = -\frac{i}{2\hbar^2} \sgn(A_\text{t}) [4 h_{\text{n},z}^2 \tilde{S}_\text{imag}(\Omega)+  
h_{\text{n},x}^2(\tilde{S}_\text{imag}(-\tilde{\omega}_\text{res} + \Omega)+\tilde{S}_\text{imag}(\tilde{\omega}_\text{res} + \Omega))]  ,\\
&& \tilde{S}_\text{imag}(\omega) \equiv  (2 i \,c_{\varepsilon}^2/\omega) \ln|\omega/\omega_l| . 
\end{eqnarray}
Note that $\Gamma_{\text{nMn}\varphi}$ is the rotating frequency induced by the low-frequency part of the $1/f$ noise.
For a driven charge qubit operated at a sweet spot $\varepsilon=0$, as considered in the main text, $h_{\text{n},z}=0$ so that terms proportional to $\tilde S(\Omega)$ or $\tilde S_\text{imag}(\Omega)$ vanish.
Since the remaining terms in $\Gamma_{\text{nMn}\varphi}$ now have opposite signs and similar magnitudes, we see that $K_{\text{nMn}\varphi}$ is small compared to $K_\text{M}$. 
If we further take $\phi = \pi/4$ as in the main text, the expressions are simplified as $\Gamma_x = \Gamma_y$ and $\Gamma_x + \Gamma_{xy} = \Gamma_z$, yielding Eq.~(5) of the main text when the initial state is $|0\rangle$.

The leading order contribution to $K_{\varphi}(t)$ is $O[\gamma^2]$: 
\begin{equation}
 K_{\varphi}(t) = 
-4 t^2(\ln(1/\omega_l t)-\gamma_E +3/2) h_{\text{n}x}^2 \frac{A_{\ell}^2}{\hbar^2\tilde{\omega}_\text{res}^2}
\begin{pmatrix}
\sin^2(\phi) & \cos(\phi)\sin(\phi) & 0 \\
\cos(\phi)\sin(\phi) & \cos^2(\phi) & 0 \\
0 & 0 & 1
\end{pmatrix} .
\end{equation}
As $\hbar d \Omega / d\varepsilon = 2 h_{\text{n}x} A_{\ell} / \hbar \tilde{\omega}_\text{res}$, we again interpret $K_{\varphi}(t)$ as the dephasing in the rotating frame, in which the eigenenergies of the dressed states are $\pm \hbar \Omega$.

For the data of the asymptotic formula shown in the main text [Fig. 1(c) and Fig. 2], we also include contributions to $K_{\varphi}$ up to $O[\gamma^2]$, and contributions to $K_\text{M}$ and $K_{\text{nMn}\varphi}$ up to $O[\gamma]$.
To capture the essence of the dynamics in a simple and illustrative way, we did not include the $O[\gamma^2]$ contributions to $K_\text{M}$ and $K_{\text{nMn}\varphi}$. 
This is justified by the fact that the lower-order contributions already capture the features of the simulation as shown in the Fig. 1(c) in the main text. 
In fact, the $O[\gamma^2]$ contributions to $K_\text{M}$ and $K_{\text{nMn}\varphi}$ are expected to have small effects.
This is contrary to $K_{\varphi}$ where the leading-order terms are $O[\gamma^2]$ and the dephasing function $I(t,\omega_1 = 0,\omega_2 = 0) \sim t^2\log(t) \gg t$ when $t$ is large, suggesting appreciable effects on the dynamics, as shown in the lower inset of Fig. 2.   


\subsection{\red{Lorentzian Noise}}
\red{
Solid-state qubit implementations contain a large number of defects that can act as charge noise sources. However, a small number of fluctuators in close proximity to a qubit could potentially dominate the noise spectrum. It is therefore interesting to consider the case of an individual fluctuator, which gives rise to random telegraph noise, whose noise spectral density ahs a well-known Lorentzian form:~\cite{doi:10.1063/1.1721637}
\begin{equation}
\label{Eq:SpectrumRTN}
\tilde{S}(\omega) =  c_{\varepsilon}^2 \frac{2 \Gamma}{\Gamma^2 + \omega^2},
\end{equation}
where $1/\Gamma$ is the switching time of the fluctuator. The corresponding time correlation function is given by 
\begin{equation}
\label{Eq:TimeCorrelationRTN}
S(t) = c_{\varepsilon}^2  e^{-\Gamma t}.
\end{equation}
In semiconductor devices, an ensemble of fluctuators with a distribution of switching times is known to give rise to noise with a $1/f$ power spectral density~\cite{PhysRevB.72.134519,2002physics...4033M}, which we investigated above, in Sec.~\ref{Sec:OneOverF}.
Below, we focus on the case of a single fluctuator. 
}

\red{
Substituting Eq.~(\ref{Eq:TimeCorrelationRTN}) in Eqs.~(\ref{Eq:ftcos})-(\ref{Eq:f0}) yields  
\begin{align}
f_\text{tcos}(t,\omega) =&  c_{\varepsilon}^2 \frac{(\Gamma^2-\omega^2) -(\Gamma^2 (1 + \Gamma t) + (-1 + \Gamma t)\omega^2)e^{- \Gamma t}\cos (\omega t) + \omega(\Gamma (2 + \Gamma t) + \omega^2 t))e^{- \Gamma t}\sin (\omega t) }{(\Gamma^2 + \omega^2)^2}, \\
f_\text{tsin}(t,\omega) =&  c_{\varepsilon}^2 \frac{2 \Gamma \omega - \omega(2 \Gamma + \Gamma^2 t + \omega^2 t) e^{- \Gamma t}\cos (\omega t) - (\Gamma^2 + \Gamma^3 t -\omega^2 + \Gamma \omega^2 t )e^{- \Gamma t}\sin (\omega t) }{(\Gamma^2 + \omega^2)^2}, \\
f_\text{t}(t) =& c_{\varepsilon}^2 \frac{1 - e^{- \Gamma t}(1+ \Gamma t)}{\Gamma^2} ,\\
f_\text{cos}(t,\omega) =& c_{\varepsilon}^2 \frac{\Gamma-\Gamma e^{- \Gamma t} \cos (\omega t) + \omega e^{- \Gamma t} \sin (\omega t)}{\Gamma^2 + \omega^2} , \\
f_\text{sin}(t,\omega) =& c_{\varepsilon}^2 \frac{\omega-\omega e^{- \Gamma t} \cos (\omega t) - \Gamma e^{- \Gamma t} \sin (\omega t)}{\Gamma^2 + \omega^2} , \\
f_\text{0}(t) =& c_{\varepsilon}^2 \frac{1 - e^{- \Gamma t}}{\Gamma}.
\end{align}
The $I(t,\omega_1,\omega_2)$ integrals then can be solved as 
\begin{equation}
\label{Eq:IntergalLorentzian}
I(t,\omega_1,\omega_2) 
= \left\{
\begin{array}{ll}
c_{\varepsilon}^2 \frac{1}{\Gamma}t + c_{\varepsilon}^2 \frac{e^{-\Gamma t}-1}{\Gamma^2} & (\omega_1 = \omega_2 = 0),  \\
c_{\varepsilon}^2  \frac{\Gamma+ i\, \omega_1}{\Gamma^2+\omega_1^2} t + c_{\varepsilon}^2 \frac{e^{-(\Gamma - i\, \omega_1)t}-1}{(\Gamma - i\, \omega_1)^2}  & (\omega_1 = - \omega_2 \neq 0),  \\
-i c_{\varepsilon}^2 \frac{e^{i\, \omega_1 t}-1}{\Gamma \omega_1} +c_{\varepsilon}^2  \frac{e^{-(\Gamma - i\, \omega_1)t}-1}{\Gamma(\Gamma - i\, \omega_1)} & (\omega_1 \neq 0 \text{ and } \omega_2 = 0)\\
-i c_{\varepsilon}^2 \frac{e^{i\, \omega_2 t}-1}{\omega_2 (\Gamma + i\,\omega_2)} + c_{\varepsilon}^2 \frac{e^{-\Gamma t}-1}{\Gamma (\Gamma + i\,\omega_2)}  & (\omega_1 = 0 \text{ and } \omega_2 \neq 0),  \\
c_{\varepsilon}^2 \frac{e^{-(\Gamma - i\, \omega_1)t}-1}{(\Gamma - i\, \omega_1)(\Gamma + i\, \omega_2)} - i c_{\varepsilon}^2 \frac{e^{-i(\omega_1+\omega_2)t}-1}{(\Gamma + i\, \omega_2)(\omega_1 + \omega_2)}  & (\text{Otherwise}) ,
\end{array}
\right.  \nonumber
\end{equation}
}

\red{
If we consider the regime $t/(2\pi) \gg 1/\omega \gg 1/\Gamma$, where $t$ is the qubit evolution time we are interested in and $\omega$ is any angular frequency relevant to the system, and drop the oscillatory terms in the full formula, then we recover our previous results for Markovian noise.
On the other hand, if we consider the regime $t/(2\pi) \gg 1/\omega \sim 1/\Gamma$ and drop the oscillatory terms in the full formula, the integrals $I(t,\omega_1,\omega_2)$ can be approximated as
 \begin{equation}
\label{Eq:IntergalLorentzianApprox}
I(t,\omega_1,\omega_2) 
\approx \left\{
\begin{array}{ll}
c_{\varepsilon}^2  \frac{\Gamma+ i\, \omega_1}{\Gamma^2+\omega_1^2} t  & (\omega_1 = - \omega_2),  \\
0 & (\text{Otherwise}) .
\end{array}
\right.  \nonumber
\end{equation}
In this case, there is not only a Markovian contribution, but also a non-Markovian-non-dephasing contribution appearing as noise-induced rotation.
}
\red{
For a system subject to noise induced by a single fluctuator or a few fluctuators, we therefore expect our protocol for improving the fidelity, as proposed in the main text, will still work, because this method was developed to address the effects of strong driving, and is not tailored to a specific power-spectral density. 
However, the effectiveness of the protocol still depends on the noise spectral density.
For example, Markovian and non-Markovian contributions to the decoherence depend differently on the noise strength at different frequencies, yielding a decoherence rate that depends on the specific form of the spectral density.
}

\section{Details of Numerical Simulations}
\label{section:simulation}
We now describe the method used for the numerical simulations.
The objective is to simulate the evolution of the ensemble average of the density matrix, $\langle \rho \rangle$, in the presence of detuning noise $\delta \varepsilon(t)$, where $\langle \cdots \rangle$ denotes the ensemble average over $\delta\varepsilon(t)$. 

The dynamics is governed by the Schr$\ddot{\text{o}}$dinger equation
\begin{equation}
\label{Eq:SchrodingerEq}
i \hbar \frac{d}{dt} |\psi(t)\rangle =[\Hamiltonian_\text{sys} + \Hamiltonian_\text{n}] |\psi(t)\rangle= [\Hamiltonian_\text{sys} + \delta \varepsilon (t) h_\text{n}] |\psi(t)\rangle .
\end{equation}
Here, $\Hamiltonian_\text{sys} = \Hamiltonian_\text{q} + \Hamiltonian_\text{ac}$ is the Hamiltonian of the system, where $\Hamiltonian_\text{q}$ describes the qubit and $\Hamiltonian_\text{ac}$ describes the ac driving, and 
$\Hamiltonian_\text{n}=h_\text{n} \delta \varepsilon(t)$ describes the effect of charge noise on the system, where $h_\text{n}$ is the noise matrix.
Using the states $\{|0\rangle = (|L\rangle-|R\rangle)/\sqrt{2},|1\rangle = (|L\rangle+|R\rangle)/\sqrt{2}\}$ as the basis,
we can express $\Hamiltonian_\text{q} = -(\epsilon/2) \sigma_x - \Delta \sigma_z$, $\Hamiltonian_\text{ac} = (A_{\varepsilon}/2)\sigma_x \cos(\omega t)$, and $h_\text{n} = -\sigma_x /2$, where the $\sigma_i$ are Pauli matrices.

The most obvious way to proceed is to first generate a set of noise realizations, $\{ \delta \varepsilon^{\alpha}(t)\}_{\alpha = 1,\ldots,\alpha_\text{max}}$, having the desired spectrum, $\tilde{S}(\omega)$, then solve the Schr$\ddot{\text{o}}$dinger equation for a given noise realization $\delta \varepsilon^{\alpha}(t)$ to obtained the final state $|\psi^{\alpha}\rangle$ and the corresponding density matrix $\rho^{\alpha} = | \psi^{\alpha}\rangle \langle \psi^{\alpha}|$.
The ensemble average of the density matrix is then given by
\begin{equation}
\langle \rho \rangle \approx \frac{1}{\alpha_\text{max}}\sum\limits_{\alpha = 1}^{\alpha_\text{max}}\rho^{\alpha}.
\end{equation}
Performing numerical simulation in this manner, however, is very time-consuming, especially when $\alpha_\text{max}$ is large.
We therefore develop a more efficient method to simulate the Schr$\ddot{\text{o}}$dinger equation over a large number of noise realizations.
We now first describe the generation of the noise realizations, and then the simulation method.

\subsection{Generation Of Noise Realizations for the Numerical Simulations}
In the simulation, a noise realization $\delta \varepsilon(t)$ is discretized into the noise sequences $\{\delta \varepsilon_1,\delta \varepsilon_2,\cdots,\delta \varepsilon_N\}$, such that $\delta \varepsilon(t) = \delta \varepsilon_k$ is constant over the time interval $t_{k-1} \leq t < t_k$ where $t_k \equiv k\, t_\text{tot}/N$, $t_\text{tot}$ is the total time of consideration, and $N$ is the total number of time segments.
We follow \cite{Kawakami18102016} to generate the noise sequences for the numerical simulation.
The method is:
\begin{enumerate}
\item Generate a Gaussian white noise sequence $\{u_1, u_2, \cdots, u_N\}$ with zero mean and unit standard deviation.
\item Compute the Fourier transform of the sequence, $\{\tilde{u}_1,\tilde{u}_2, \cdots, \tilde{u}_N\}$, where
\begin{equation}
\tilde{u}_m = \sum\limits_{n=1}^N u_n e^{-i 2 \pi \frac{(n-1)(m-1) }{N}}. 
\end{equation}
\item Let $\tilde S(\omega)$ represent the desired noise power spectrum. Scale the white noise in the Fourier space by taking the product $\tilde{u}_m \sqrt{S^{\prime}_m}$, where $S^{\prime}_m = \tilde S\left( 2\pi (m-1)/t_\text{tot} \right) /\Delta t$, and $\Delta t = t_\text{tot}/N$. Then perform the inverse Fourier transformation to obtain the noise realization in the time domain:  
\begin{equation}
\label{Eq:NoiseSeries}
\delta \varepsilon_k = \frac{1}{N} \sum\limits_{m=1}^N \tilde{u}_m \sqrt{S^{\prime}_m} e^{i 2 \pi \frac{(m-1)(k-1) }{N}}.
\end{equation}
Here, we assume $1/f$ noise such that
\begin{equation}
 S^{\prime}_m = \\ 
 c_{\varepsilon}^2 
\left\{
\begin{array}{ll}
N/(m-1) & m = m_l,2,\ldots,N/2+1\\
N/(N - m+1) & m = N/2+2,\ldots,N-m_l+2
\end{array} 
\right. \nonumber , 
\end{equation}
corresponding to the continuous noise spectrum
\begin{equation}
\label{Eq:Spectrum One-Over-f}
\tilde{S}(\omega) = c_{\varepsilon}^2 \left\{
  \begin{array}{cl}
    \frac{2 \pi}{|\omega|} & : \omega_l \leq |\omega | \leq \omega_h\\
    0 & : \text{otherwise} 
  \end{array}
\right. ,
\end{equation}
where the low angular frequency cutoff is $\omega_l = 2 \pi (m_l-1) / t_\text{tot}$ and the high angular frequency cutoff is $\omega_h = \pi N / t_\text{tot}$.
\end{enumerate}

\subsection{Method of Simulation}
We wish to solve the Schr$\ddot{\text{o}}$dinger equation, Eq.~(\ref{Eq:SchrodingerEq}), numerically for a given noise realization $\delta \varepsilon (t)$ defined as in Eq.~(\ref{Eq:NoiseSeries}).
Let $U_0(t)$ be the evolution operation satisfying 
\begin{equation}
i \hbar \frac{d}{dt} U_0(t) = \Hamiltonian_\text{sys} U_0(t),
\end{equation}
with initial condition $U_0(t=0) = I$.
We transform the Schr$\ddot{\text{o}}$dinger equation into the interaction picture, defining $|\psi^I(t)\rangle = U_0(t)^{\dagger} |\psi(t)\rangle$ and $h_\text{n}^I(t) = U_0(t)^{\dagger} h_\text{n}(t) U_0(t)$, so that 
\begin{equation}
i \hbar \frac{d}{dt} |\psi^I(t)\rangle = \delta \varepsilon (t)  h_\text{n}^I(t) |\psi^I(t)\rangle .
\end{equation} 

To include the cases where gate times are in between $t_k$ and $t_{k-1}$, we define the evolution operator $U^I(t,t_{k-1})$ as $|\psi^I(t)\rangle = U^I(t,t_{k-1}) |\psi^I(t_{k-1})\rangle$ for $t_k \geq t \geq t_{k-1}$, which can be expanded perturbatively as
\begin{eqnarray}
U^I(t,t_{k-1}) 
& =& I + \sum\limits_{r = 1}^{\infty} (\delta \varepsilon_k)^r W_r(t,t_{k-1}), \nonumber
\end{eqnarray}
where $W_r(t,t_{k-1})$'s are time-ordered integrals, 
\begin{eqnarray}
W_1(t,t_{k-1}) &\equiv& \left(-\frac{i}{\hbar}\right) \int_{t_{k-1}}^t d\tau_1 \,  h_\text{n}^I(\tau_1), \\
W_r(t,t_{k-1}) &\equiv& \left(-\frac{i}{\hbar}\right)^r \int_{t_{k-1}}^t d\tau_1 \cdots \int_{t_{k-1}}^{\tau_{r-1}} d\tau_r \,  h_\text{n}^I(\tau_1) \cdots h_\text{n}^I(\tau_r), \; \text{ for } r = 2,3,\cdots.
\end{eqnarray}
Truncating the series after $r=r_\text{max}$, one can approximate the evolution operator as
\begin{equation}
\label{Eq:UIt_tk}
U^I(t,t_{k-1}) = I + \delta \varepsilon_k W_1(t,t_{k-1}) + (\delta \varepsilon_k)^2 W_2(t,t_{k-1})+ \cdots + (\delta \varepsilon_k)^{r_\text{max}} W_{r_\text{max}}(t,t_{k-1})  .
\end{equation}
where $\{W_r(t,t_{k-1})\}$ can be calculated either analytically or numerically.
The evolution operator $U^I(t,0)$ then can be expressed as
\begin{equation}
\label{Eq:UIt}
U^I(t,0) = 
U^I(t,t_{k-1})U^I(t_{k-1},t_{k-2})\cdots U^I(t_{1},0) .
\end{equation}
The solution of the Schr$\ddot{\text{o}}$dinger equation in the lab frame can then be expressed as $|\psi(t)\rangle = U(t)|\psi(0)\rangle $, where $U(t) = U_0(t) U^I(t,0)$.

The strength of this method is that $W_r(t,t_{k-1})$ does not depend on delta $\varepsilon(t)$, so we only need to perform simulations once, to determine $\{W_r(t,t_{k-1})\}_{r=1,\ldots,r_\text{max}}$ and $\{W_r(t_{l},t_{l-1})\}_{r=1,\ldots,r_\text{max}}$ for $l = 1,\ldots,k-1$, numerically.
After computing these matrices, the calculation of $|\psi(t)\rangle$ for all the generated noise sequences directly follows Eq.~(\ref{Eq:UIt_tk}) and Eq.~(\ref{Eq:UIt}), with no further simulations required.
This allows efficient simulation of the system, even with a large number of noise realizations.

In this work, we take $r_\text{max} = 6$, $\alpha_\text{max}=100,000$, and compute $\{W_r\}$ numerically. 
We have checked the validity of this simplified method by comparing our results to those of full simulations with parameters the same as the simulation shown in Fig. 1(b) and Fig. 1(c) for several different noise sequences, and find the deviations between the resulting density matrices are smaller than $10^{-7}$ for the time period considered.
We also study the statistics of the simulated fidelities by dividing the simulation into $10$ trials, each with $\alpha_\text{max}=10,000$, and compute the ratio of the standard deviation of these $10$ trials, $\sigma_F$, to the average infidelity $\sigma_F/(1-F)$. 
For all the parameters considered in the main text, this ratio is around $10^{-2}$, showing that the reported infidelity is accurate to at least two significant figures.


\section{High-fidelity $R_{\theta}(\phi=\pi/4)$ Gates}
\begin{figure}[t]
\includegraphics[width=4in]{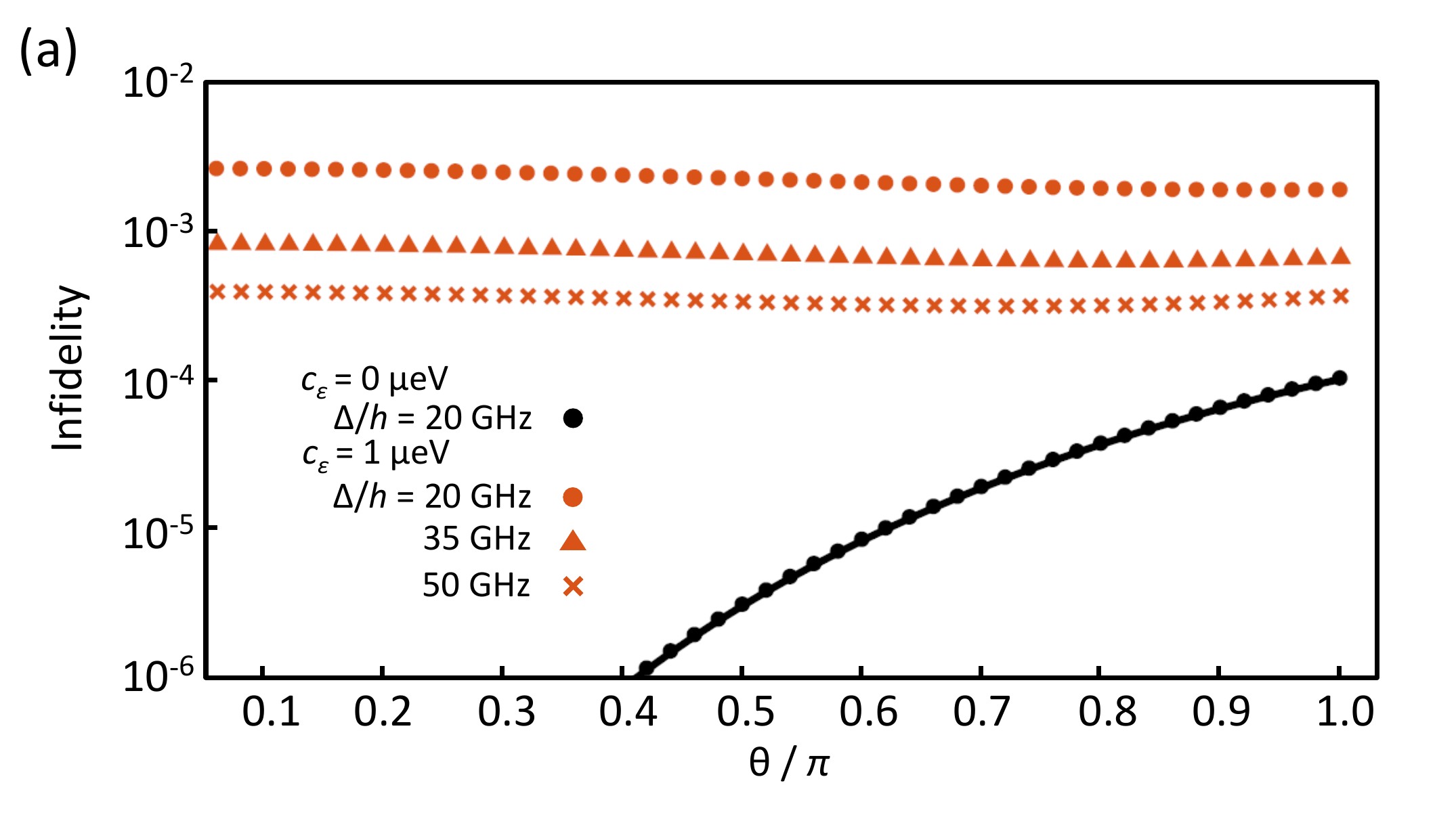}
\caption{
\label{Fig:FigS1}
\red{Characterization of fidelity of gates with different rotation angles.}
Infidelity of an $R_{\theta}(\phi=\pi/4)$ gate performed on a strongly driven charge qubit in the presence of detuning noise while operated in the dip with $N = (2 \theta \tilde{\omega}_\text{res})/(\pi \Omega) = 10$. Note that all infidelities plotted here are $< 0.01$. The qubit is operated at the sweet spot where the detuning $\varepsilon = 0$, and the different curves show the results for different tunnel couplings from $\Delta/h = 20\, \units{GHz}$ to $50\, \units{GHz}$, as indicated. The noise spectrum is $1/f$, $\tilde{S} = c_{\varepsilon}^2 2 \pi/\omega$, with low (high) angular frequency cutoff $\omega_l/2\pi = 1 \, \units{Hz}$ ($\omega_h/2\pi = 256 \, \units{GHz}$), where the low frequency part (frequency smaller than $0.3 \, \units{MHz}$) is approximated as quasi-static noise in the simulations. The figure shows simulations with noise amplitude $c_{\varepsilon} = 0\, \units{\mu eV}$ ($\Delta/h = 20\, \units{GHz}$, black dots), and $c_{\varepsilon} = 1\, \units{\mu eV}$ ($\sigma_{\varepsilon} = 6.36\, \units{\mu eV}$) for $\Delta/h = 20 \, \units{GHz}$  (orange dots), $\Delta/h = 35 \,\units{GHz}$ (triangles) and  $\Delta/h = 50 \, \units{GHz}$ ($\times$ markers), and also analytic results [Eq.~(\ref{Eq:IntrinsicInfidelityPi4})] for $c_{\varepsilon} = 0\, \units{\mu eV}$ and $\Delta/h = 20 \,\units{GHz}$ (black lines). The simulation shows that as $\Delta$ and $A_{\varepsilon}$ increase, the gate time decreases and the infidelities approach the intrinsic infidelity limit, corresponding to $c_{\varepsilon} = 0\, \units{\mu eV}$. Note that when $\Delta/h \gtrsim 35 \,\units{GHz}$, these gate fidelities are all greater then $99.9\%$.
}
\end{figure}

\label{section:RthetaGates}
According to Eq.~(\ref{Eq:FidelityAt}), to $O[\gamma^3]$, the intrinsic infidelity of a resonantly driven $R_{\theta}(\phi=\pi/4)$ gate performed on a charge qubit in the absence of noise can be expressed as
\begin{equation}
1- F_{\theta}(\phi=\pi/4) =  2 \gamma ^2 [1 - \cos(2 \theta \tilde{\omega}_\text{res}/\Omega )] + 4 \gamma^3 \sin(\theta)\sin(2 \theta \tilde{\omega}_\text{res}/\Omega).
\end{equation}
For any rotation angle $\theta$, the intrinsic infidelity vanishes whenever $N \equiv (2\tilde{\omega}_\text{res} \theta) /(\Omega \pi)$ is an even integer, producing dips like those observed in Fig.~3 of the main text. 
However, in the presence of the noise, the dips are suppressed and the infidelity increases.
This effect can be reduced by increasing the tunnel coupling $\Delta$ and the driving amplitude $A_{\varepsilon}$ simultaneously while keeping the ratio $A_{\varepsilon}/\Delta$ fixed, so that the system remains in a dip. 
This is because the strong-driving effects only depend on the ratio of the driving amplitude and the tunnel coupling $A_{\varepsilon}/\Delta$; however, by reducing the gate time ($t_g \propto A_{\epsilon}^{-1}$), we can reduce the effects of charge noise, thus improving the fidelity.
This method was applied to the special case of $R_{\pi}(\phi=\pi/4)$ gates in the main text.
To demonstrate that the method works for any $\theta$, we now apply it to $R_{\theta}(\phi=\pi/4)$ gates for $0 < \theta < \pi$.

In Fig.~\ref{Fig:FigS1}, we compare numerical results for the intrinsic infidelities ($c_{\varepsilon} = 0 \,\units{\mu eV}$) of $R_{\theta}(\phi=\pi/4)$ gates with infidelities in the presence of $1/f$ detuning charge noise with $c_{\varepsilon} = 1\, \units{\mu eV}$ ($\sigma_{\varepsilon} =6.36\, \units{\mu eV}$), operated in the dip corresponding to $N = 10$. 
As usual, the system is operated at the sweet spot ($\varepsilon=0$), and we consider the tunnel couplings $\Delta/h = 20, \, 35, \, 50 \, \units{GHz}$. 
Since the intrinsic infidelity vanishes up to $O[\gamma^3]$ when operating in a dip, its leading order contribution, from Eq.~(\ref{Eq:Fidelity4}), is given by
\begin{equation}
\label{Eq:IntrinsicInfidelityPi4}
1-F_{\theta}(\pi/4) = 16 \gamma^4  \sin^2(\theta/2),
\end{equation}
which increases as $\theta$ increases, and serves as the lower bound of the infidelity in the presence noise.
When $c_{\varepsilon} = 1\, \units{\mu eV}$ ($\sigma_{\varepsilon} =6.36\, \units{\mu eV}$), decoherence dominates the infidelity, which is then much larger than the intrinsic infidelity.
By increasing $A_{\varepsilon}$ and $\Delta$ while keeping their ratio fixed, as discussed above, the gate time can be reduced, causing the infidelity to approach the intrinsic infidelity limit, as shown in Fig.~\ref{Fig:FigS1}.
This demonstrate that our method for improving the fidelity is applicable to a generic $R_{\theta}(\phi=\pi/4)$ rotation, not just for the case $\theta = \pi$ that is discussed in the main text.

\end{widetext}




\end{document}